\documentclass[12pt]{article}

\usepackage{algorithm}
\usepackage{algpseudocode}
\usepackage{amsmath}
\usepackage{amsfonts}
\usepackage{amssymb}
\usepackage{svg}
\usepackage{graphicx}
\usepackage{subcaption}
\usepackage{enumerate}
\usepackage{geometry}
\usepackage[colorlinks=true, allcolors=blue]{hyperref}
\usepackage[round]{natbib}
\usepackage{xcolor}
\usepackage{esvect}
\usepackage{caption}
\usepackage{setspace}
\usepackage{caption}
\usepackage{xcolor}
\usepackage[english]{babel}
\usepackage{amsthm}
\usepackage[toc,page]{appendix}
\usepackage[textsize=tiny, textwidth=2cm, disable]{todonotes} 
\usepackage{booktabs,threeparttable}

\theoremstyle{plain}
\newtheorem{theorem}{Theorem}[section]

\newtheorem{proposition}{Proposition}[section]

\newtheorem{lemma}[theorem]{Lemma}
\newtheorem{assumption}{Assumption}
\newtheorem{identificationassumption}{Assumption}

\theoremstyle{definition}

\newcommand{\cov}{\operatorname{Cov}}

\newcommand{\Vy}{\boldsymbol{y}}
\newcommand{\Vx}{\boldsymbol{x}}
\newcommand{\Va}{\boldsymbol{\alpha}}
\newcommand{\Vbeta}{\boldsymbol{\beta}}
\newcommand{\Vgamma}{\boldsymbol{\gamma}}
\newcommand{\Vw}{\boldsymbol{w}}

\newcommand{\Vep}{\boldsymbol{\ep}}

\newcommand{\Vtheta}{\boldsymbol{\theta}}
\newcommand{\TZ}{\widetilde {Z}}

\newcommand{\TVy}{\widetilde {\Vy}}

\newcommand{\Vc}{\boldsymbol {c}}
\newcommand{\TX}{\widetilde {X}}
\newcommand{\TVep}{\widetilde \Vep}
\newcommand{\TVbeta}{\widetilde {\Vbeta}}

\newcommand{\EE}{\mathbb E}
\newcommand{\RR}{\mathbb R}

\newcommand{\noi}{\noindent}

\newcommand{\ep}{\varepsilon}

\title{\textbf{Network-Adaptive Instrumental Variables}}

\author{
Yi Niu\thanks{Department of Economics, University of Pennsylvania. Email: \texttt{n1y1@sas.upenn.edu}}
\and
Zhengyan Xu\thanks{Department of Economics, University of Pennsylvania. Email: \texttt{xzy24@sas.upenn.edu}}
}
\date{November 2026}

\begin{document}
\onehalfspacing

\maketitle

\begingroup
\renewcommand{\thefootnote}{}
\footnotetext{We thank Karun Adusumilli, Xu Cheng, Haobai Guo, Wayne Gao, Ming Li, Xun Tang, and Zhiheng You, as well as participants at the University of Pennsylvania Econometrics Lunch, for helpful comments.}
\addtocounter{footnote}{0}
\endgroup

\begin{abstract}
\noi Three mechanisms generate endogeneity in regressions under interference. We address them by constructing network-adaptive instrumental variables (NIV) for a spatial autoregressive (SAR) model with endogenous covariates and unknown network weights. Under network-conditional exclusion and information equivalence, the resulting instruments yield valid moments without knowing
the true network. We establish joint identification of the structural mean coefficients and the network parameters, providing two sufficient routes. Then under additional covariance restrictions, we identify the disturbance spillover coefficient and the innovation variance. We develop NIV-GMM estimation, establish consistency and asymptotic normality, and give feasible inference and a test for mean network spillovers. Monte Carlo simulations examine their finite-sample performance.

\end{abstract}

\newpage

\section{Introduction}

Endogeneity is a key concern in regression analysis, and it becomes especially consequential in models with interference, peer effects, or network spillovers \citep{Manski1993,LEE2007,BRAMOULLE200941,Graham2018}. Three mechanisms interact in such settings. First, the covariates can be correlated with unobserved determinants of the outcome. Second, network link formation or link strength may depend on endogenous characteristics \citep{JacksonWolinsky1996,Graham2017,dePaula2017}. Third, outcomes, covariates, and shocks may propagate through the resulting network. In the presence of these mechanisms, an instrument that is valid in the absence of interference is likely to fail.

The above mechanisms arise in many empirical settings. For example, infrastructure and human-capital investment can be endogenous in a regression of city-level economic growth. Their effects may also propagate across cities, while the strength of intercity exposure may depend on economic or demographic similarity. The empirical problem is to construct moments that remain valid in this case even if the true network between cities is unknown.

The literature addressing endogeneity in such settings can be broadly categorized into two strands. One strand builds on traditional instrumental variables (IV) methods, and the other strand relies on spatial and network models. Traditional IV methods address endogenous regressors without explicitly modeling the network interference structure \citep{Wooldridge2010}. Spatial and network literature instead models interference directly, commonly through a spatial autoregressive (SAR) or linear-in-means specification. The latter literature identifies important reflection and correlated-effects problems \citep{Manski1993,LEE2007,BRAMOULLE200941} and develops IV and GMM estimators for spatially lagged outcomes and covariates \citep{Kelejian1998,KelejianPruchaYuzefovich2004,LIU201099,LiuLee2013}. However, much of this work treats the network weighting matrix as exogenously given when the instruments are constructed and the moment conditions are formed. Several recent papers relax that restriction. For example, \citet{QuLee2015} develops estimation for SAR models with an endogenous spatial weight matrix, and \citet{LeeLiebermanPirasRossi2025} studies identification and inference with endogenous weights. Nevertheless, relaxing one restriction is often at the cost of compromising on others, leaving important sources of endogeneity unaddressed. Handling endogenous regressors in an SAR model requires carefully tailored IV constructions and instrument selection. The problem becomes even harder when the network weights are endogenous, or the shocks involve autoregressive structures, and instruments can be numerous \citep{Kelejian1998,KelejianPrucha2010,LIU201099,LiuLee2013}.

This paper studies a spatial autoregressive model with endogenous covariates and unknown, potentially endogenous network weights generated by observed distances, targeting at all three mechanisms discussed at the beginning. We construct network-adaptive instrumental variables (NIV) by residualizing candidate instruments with respect to available network information and applying powers of the candidate network weighting matrix. Under the network-conditional exclusion assumption and information equivalence, this construction yields valid moment conditions without knowing the true network weights. We establish sufficient conditions for joint identification of the structural mean coefficients and the group-specific network parameters, building upon a global separation condition that rules out compensating changes in coefficients and network weights. We also provide separate conditions for identifying the disturbance parameters. For estimation, we develop an NIV-GMM estimator and establish consistency and asymptotic normality under additional regularity conditions. We construct feasible confidence intervals and a test for mean network spillovers based on fixed moment restrictions, and examine finite-sample performance through Monte Carlo simulations.

The rest of the paper is organized as follows. Section \ref{section: model setup} sets up the model. Section \ref{section: NIV} constructs the network-adaptive instruments. Section \ref{section: identification} establishes joint identification of the network parameters and structural mean coefficients through a high-level global separation condition and two alternative sufficient routes, and then treats the disturbance parameters. Section \ref{section: estimation} presents NIV--GMM estimation, its consistency and asymptotic normality conditions, feasible inference, and the test for mean network spillovers. Section \ref{section: simulation} reports Monte Carlo evidence on finite-sample performance. Section \ref{section: conclusion} concludes.

\section{Model Setup}\label{section: model setup}
\noi Consider the linear regression:
\begin{align}\label{model:naive}
           y_{i}  & = \Vx_i^{\top}\Vbeta_{0}+ \ep_i^*
\end{align}
where $y_i$ is the outcome and $\Vx_i = (x_{1i},\dots, x_{ki})^{\top}$ is a $k$-dimensional vector of covariates for unit $i\in \{1,2,\dots,N\}$. The OLS estimator for $\Vbeta_{0}\in \mathbb R^{k}$ is inconsistent if $\Vx_i$ is correlated with $\ep_i^*$. An instrumental variable strategy is natural. However, a naive instrumental variable may fail when the reduced-form disturbance $\ep_i^*$ contains spillovers propagating through a network. We therefore consider the following true model:
\begin{subequations}\label{Theory: True Model}
\begin{align}
    y_i& =  \Vx_i^{\top}\Vbeta_{10}+ \text{Spill}_{i,r_i}(\theta_{0r_i},\rho_0,\Vbeta_{20},\delta_0) + \ep_i,\\
    \text{Spill}_{i,r_i}(\theta_{0r_i},\rho_0,\Vbeta_{20},\delta_0) & = \rho_0y_{-i,r_i}^{\text{net}}(\theta_{0r_i}) +\Vx_{-i,r_i}^{\text{net}}(\theta_{0r_i})^{\top} \Vbeta_{20}  + \alpha_{0r_i} + \delta_0 \ep_{-i,r_i}^{\text{net}}(\theta_{0r_i}).
\end{align}
\end{subequations}
\noi Here $r_i \in \{1,\dots,\bar r\}$ denotes the group containing unit $i$. The group membership is exogenously fixed according to the sampling design, administrative boundary, or field-specific knowledge. Variables with superscript ``net'' are aggregations of exposure to other units, denoted by $-i$, in the same group through a network. This model belongs to the well-known class of SAR models, where we model the interference within each prespecified group.

\noi  The network is modeled through observed dyadic distances. Let 
$$
d_{ij}\equiv d(\Vw_i,\Vw_j)\in\RR_+,\qquad d_{ij}=d_{ji}.
$$
The vector $\Vw_i$ contains observed characteristics relevant for homophily or economic proximity. For units in group $r$, define the off-diagonal weight 
$$
g_{r,ij}(\theta_r)\equiv a_{r,n}K_{\theta_r}(d_{ij}),\qquad i\neq j,
$$
and set $g_{r,ii}(\theta_r)=0$. The factor $a_{r,n}>0$ is known and useful for asymptotic analysis. For now, we consider $a_{r,n}=1$ and a fixed finite network unless otherwise specified. The kernel $K_{\theta_r}:\RR_+\to[0,1]$ is parametric, and $\theta_r$ is a group-specific decay parameter. The baseline exponential-quadratic kernel is 
$$
K_{\theta_r}(d_{ij})=\exp(-\theta_r d_{ij}),\qquad d_{ij}=\|\Vw_i-\Vw_j\|^2.
$$
Larger values of $\theta_r$ produce more localized exposure. With $\mathcal{J}_{r_i}\equiv\{j\neq i:r_j=r_i\}$, define 
$$
y_{-i,r_i}^{\mathrm{net}}(\theta_{0r_i})=\sum_{j\in\mathcal{J}_{r_i}}g_{r_i,ij}(\theta_{0r_i})y_j,\quad \Vx_{-i,r_i}^{\mathrm{net}}(\theta_{0r_i})=\sum_{j\in\mathcal{J}_{r_i}}g_{r_i,ij}(\theta_{0r_i})\Vx_j,
$$
$$
\ep_{-i,r_i}^{\mathrm{net}}(\theta_{0r_i})=\sum_{j\in\mathcal{J}_{r_i}}g_{r_i,ij}(\theta_{0r_i})\ep_j.
$$
The scalar $\alpha_{0r_i}$ is an unobserved group fixed effect.

\noi Write the model in matrix form. Let $n_r$ denote the size of group $r$, write $N=\sum_{r=1}^{\bar r}n_r$ for the total sample size, let $\Va_{0r}=\alpha_{0r}\boldsymbol 1_{n_r}$, and define $\mathcal I_r=\{i:r_i=r\}$. For an ordering $\mathcal I_r=\{i_1,\ldots,i_{n_r}\}$, let $\Vy_r=(y_{i_1},\ldots,y_{i_{n_r}})^{\top}$, let $X_r = (\Vx_{i_1},\Vx_{i_{2}},\dots,\Vx_{{i_{n_r}}})^{\top}$, and let $\Vep_r=(\ep_{i_1},\ldots,\ep_{i_{n_r}})^{\top}$. The symmetric matrix $G_r(\theta_{0r})\in\RR^{n_r\times n_r}$ has $(h,l)$ entry $g_{r,i_hi_l}(\theta_{0r})$. Equation \eqref{Theory: True Model} becomes
\begin{equation}\label{Theory: True Model Matrix Form}
\Vy_r = \rho_0 G_r(\theta_{0r})\Vy_r +  X_r\Vbeta_{10} +G_r(\theta_{0r})X_r \Vbeta_{20}   + \Va_{0r}+\delta_0 G_r(\theta_{0r})\Vep_r + \Vep_r,
\end{equation}
\noi for $r=1,\dots,\bar r$. Within a group, the outcome, covariate, and disturbance spillovers propagate through the same network $G_r(\theta_{0r})$. We maintain that $I_{n_r}-\rho_0G_r(\theta_{0r})$ is nonsingular almost surely for every group $r$, so the structural model has a unique reduced-form solution; the stronger bound $|\rho_0|\,\|G_r(\theta_{0r})\|_{\mathrm{op}}<1$ is sufficient.

\noi Following \citet{LIU201099}, define the within-group projector $J_r\equiv I_{n_r}-n_r^{-1}\boldsymbol 1_{n_r}\boldsymbol 1_{n_r}^{\top}$. Pre-multiplication by $J_r$ removes $\Va_{0r}$ and gives
\begin{equation}\label{Model: within-differenced}
        J_r \Vy_r = \rho_0 J_r G_{0r}\Vy_r+ J_r X_r \Vbeta_{10}+  J_rG_{0r}X_r\Vbeta_{20}  +  \delta_0 J_r G_{0r}\Vep_r + J_r \Vep_r,
\end{equation}
where $G_{0r} = G_r(\theta_{0r})$.

Stack the group-specific vectors and matrices conformably. Let $G_0=\operatorname{diag}\{G_{01},\ldots,G_{0\bar r}\}$ and $J=\operatorname{diag}\{J_1,\ldots,J_{\bar r}\}$. The full model is
\begin{equation}\label{model:full}
    J \Vy = \rho_0 J G_0 \Vy + JX\Vbeta_{10} +  J G_0 X \Vbeta_{20}  +  \delta_0 J G_0 \Vep + J \Vep.
\end{equation}

Let $\TVy=J\Vy$, define
\[
\TX(\Vtheta)=\big(JG(\Vtheta)\Vy,\ JX,\ JG(\Vtheta)X\big),
\qquad
\TVbeta_0=(\rho_0,\Vbeta_{10}^{\top},\Vbeta_{20}^{\top})^{\top},
\qquad
\TVep(\delta,\Vtheta)=(J+\delta JG(\Vtheta))\Vep.
\]
With $\Vtheta_0=(\theta_{01},\ldots,\theta_{0\bar r})^{\top}$, write $\TX=\TX(\Vtheta_0)$ and $\TVep=\TVep(\delta_0,\Vtheta_0)$, then the model is in a compact form:
\begin{equation}\label{model:compact}
    \TVy = \TX \TVbeta_0 + \TVep.
\end{equation}

In the naive model \eqref{model:naive}, omitted network spillover terms are absorbed into $\ep_i^*$. A candidate instrument correlated with $X$ may then be correlated with $\ep_i^*$ through peer outcomes which involve self covariates. The consequence is the failure of the naive modeling and traditional instruments.

Standard SAR instrument matrices are constructed from a known weighting matrix \citep{LIU201099}. Here the matrix depends on the unknown $\Vtheta_0$. The next section constructs NIV moments that adapt to each candidate $\Vtheta$ and address endogeneity in both $G_0\Vy$ and $X$.

\section{Network-Adaptive Instrumental Variables}\label{section: NIV}

Let $Z\in\RR^{N\times k}$ contain $k$ candidate instruments that are relevant for $X$. For each $\Vtheta\in\Theta$, let $G(\Vtheta)$ denote the block-diagonal matrix generated by the observed distances and the candidate kernel parameters.

\begin{assumption}[Network-conditional exclusion]\label{assumption: conditional exogeneity}
The structural disturbance satisfies $$\EE[\Vep\mid Z,G_0]=\EE[\Vep\mid G_0].$$
\end{assumption}

Assumption \ref{assumption: conditional exogeneity} allows $Z$ and $\Vep$ to be dependent, but only through the network. After conditioning on the true network weighting matrix that determines exposure, the candidate instruments contain no additional information about the structural disturbance or any unobserved determinants of the outcome, at the first moment.
Note that neither the conditional expectation $\EE[\Vep\mid Z,G_0]$ nor the expectation of products such as $\EE[Z^{\top}G_0\Vep]$ needs to be zero. Such unconditional orthogonality would be unrealistically strong when relevant economic controls are omitted.

The information-equivalence condition stated in Assumption \ref{assumption: Kernel Regularity} below implies
\[
\sigma\{G(\Vtheta)\}=\sigma(G_0)
\quad\text{for every }\Vtheta\in\Theta.
\]
Consequently, the conditional mean
\[
\mu_G\equiv\EE[Z\mid G_0]=\EE[Z\mid G(\Vtheta)]
\]
is the same random matrix for every candidate value of $\Vtheta$. Let $R_Z\equiv Z-\mu_G$ denote the residualized candidate instrument matrix, then the set of valid NIV matrices is
$$\mathcal Z_{\Vtheta}=\{R_Z,\ G(\Vtheta)R_Z,\ G(\Vtheta)^2R_Z,\ldots\},$$
as shown in Proposition \ref{Proposition: Valid Moment Condition}. NIV first removes the component of $Z$ predictable from the network information and then uses candidate-network powers to generate additional orthogonal directions. A similar construction has been used in the SAR literature \citep{Kelejian1998,KelejianPrucha2010,LIU201099,LiuLee2013} but requires known network.

For a fixed highest network power $q\ge0$, the NIV matrix is
\begin{equation}\label{equation: IV p choice}
\TZ_{\Vtheta}=\big(R_Z,\;G(\Vtheta)R_Z,\;\ldots,\;G(\Vtheta)^qR_Z\big).
\end{equation}
We write the NIV matrix under the true parameter as $\TZ=\TZ_{\Vtheta_0}$. The structural mean equation contains $p=2k+1$ coefficients. The joint identification of $\Vtheta_0$ and the $p$ coefficients uses $l=(q+1)k \ge p+\bar{r}$ columns. Linear dependence can reduce the effective number of moments, so the rank condition is stated separately below. $\mathcal Z_{\Vtheta}$ offers more moment equations than structural mean parameters. Instrument selection with many network instruments is discussed in \citet{LiuLee2013}.

\section{Identification}\label{section: identification}

The identification has two steps. The first step uses the NIV moments to identify the decay parameters and the coefficients $(\Vtheta_0,\TVbeta_0)$, where
$$
\TVbeta_0 \equiv (\rho_0,\Vbeta_{10}^{\top},\Vbeta_{20}^{\top})^{\top}\in \RR^{2k+1}.
$$
The second step uses the covariance of the identified structural residual to identify the disturbance parameters $\delta_0$ and $\sigma_{\ep}^{2}$.

\subsection{Kernel regularity, validity, and relevance}

We first specify regularity conditions for the distance kernel that generates the network weighting matrix.

\begin{assumption}[Distance-kernel regularity]
\label{assumption: Kernel Regularity}
Let $\Theta\subseteq\prod_{r=1}^{\bar r}\Theta_r$ be compact. For each $\Vtheta\in\Theta$ and each group $r$, the kernel $K_{\theta_r}:\RR_+\to[0,1]$ satisfies:
\begin{enumerate}[(i)]
\item $K_{\theta_r}(d)$ is Borel measurable in $d$ and continuously differentiable in $(\theta_r,d)$ on the support of the off-diagonal distances. Moreover, for every nondegenerate distance $d>0$, the mapping $\theta_r\mapsto K_{\theta_r}(d)$ is strictly monotone.

\item For every $\theta_r,\theta_r' \in \Theta_r$, the mappings $d\mapsto K_{\theta_r}(d)$ and $d\mapsto K_{\theta_r'}(d)$ are one-to-one on the support of off-diagonal distances and are Borel measurable transformations of each other.

\item For any $\theta_r'\neq\theta_r''$, there is no constant $a>0$ such that
$$
K_{\theta_r'}(d)=aK_{\theta_r''}(d)
\qquad\text{for all off-diagonal distances }d\text{ in the support}.
$$
\end{enumerate}
\end{assumption}

The second regularity condition in this assumption gives us the information equivalence. Hence, for every $\Vtheta,\Vtheta'\in\Theta$, conditioning on $G(\Vtheta)$ is equivalent to conditioning on $G(\Vtheta')$. For the Gaussian distance kernel $K_{\theta_r}(d)=\exp(-\theta_r d)$, positive decay parameters ensure information equivalence. In particular,
$$
K_{\theta_r}(d)=K_{\theta_{0r}}(d)^{\theta_r/\theta_{0r}},
$$
so, for every $\Vtheta,\Vtheta_0\in\Theta$ with positive components, $G(\Vtheta)$ and $G(\Vtheta_0)$ are one-to-one measurable transformations of each other block by block. Nonproportionality in (iii) additionally requires at least two distinct off-diagonal distances in each block's distance support. For the Gaussian kernel, the ratio of two kernels is $\exp\{-(\theta_r'-\theta_r'')d\}$, which cannot be constant at two distinct distances when $\theta_r'\ne\theta_r''$.

\begin{proposition}[Validity of NIV moments]\label{Proposition: Valid Moment Condition}
Fix a nonnegative integer $q$ and define 
$$
\TZ_{\Vtheta}
\equiv
\big(Z-\mu_{G(\Vtheta)},\;G(\Vtheta)(Z-\mu_{G(\Vtheta)}),\;\ldots,\;G(\Vtheta)^q(Z-\mu_{G(\Vtheta)})\big).
$$
Under Assumptions \ref{assumption: conditional exogeneity} and \ref{assumption: Kernel Regularity}, for every $\Vtheta\in\Theta$,
$$
\EE\!\left[\TZ_{\Vtheta}^{\top}\TVep\right]=\boldsymbol 0.
$$
where $\TVep=(J+\delta_0JG(\Vtheta_0))\Vep$.
\end{proposition}

The proof is given in Appendix~\ref{Appendix: proof of NIV validity}.

For the identification arguments below, for any $\Vtheta,\Vtheta'
\in\Theta$, define the outcome component $\Vc_{\Vtheta}$ and covariates component $M_{\Vtheta}(\Vtheta')$ as
$$
\Vc_{\Vtheta}\equiv \EE\!\left[\frac{1}{N}\TZ_{\Vtheta}^{\top}\TVy\right],\quad \text{and} \quad 
M_{\Vtheta}(\Vtheta')\equiv \EE\!\left[\frac{1}{N}\TZ_{\Vtheta}^{\top}\TX(\Vtheta')\right].
$$
Additionally, define the self component $A_{\Vtheta}$, peer component $B_{\Vtheta}(\Vtheta')$, and true peer component coefficients $\Vgamma_0$ as
$$
A_{\Vtheta}\equiv \EE\!\left[\frac{1}{N}\TZ_{\Vtheta}^{\top}JX\right],\quad
B_{\Vtheta}(\Vtheta')\equiv \EE\!\left[\frac{1}{N}\TZ_{\Vtheta}^{\top}(JG(\Vtheta')\Vy,JG(\Vtheta')X)\right],\quad \Vgamma_0\equiv (\rho_0,\Vbeta_{20}^{\top})^{\top}\in\RR^{k+1}.
$$

Note that the terms with subscript $\Vtheta$ have the NIV $\TZ_{\Vtheta}$ constructed using $G(\Vtheta)$, and those presented as a function of $\Vtheta'$ involves spillover components under network $G(\Vtheta')$.

\begin{assumption}[Rank conditions]\label{assumption: rank conditions} For every $\Vtheta \in \Theta$, 
\begin{enumerate}[(i)]
\item The set $\mathcal{Z}_{\Vtheta}$ contains at least $2k + 1 + \bar{r}$ linearly independent columns. 
\label{assumption: Network Instruments: Construction}

\item $M_{\Vtheta}(\Vtheta)$ has full column rank $2k+1$.
\label{assumption: full rank}
\end{enumerate}
\end{assumption}

The number of instrument columns alone does not establish either part of Assumption~\ref{assumption: rank conditions}; relevance must remain after residualization and the within transformation.

\subsection{Global separation of the NIV moments}

Let $M(\Vtheta)\equiv M_{\Vtheta}(\Vtheta)$ and define
\begin{align}
 m(\TVbeta,\Vtheta)
 &=\Vc_{\Vtheta}-M(\Vtheta)\TVbeta,\label{eq:population-niv-moment}\\
 d(\Vtheta)
 &=\EE\!\left[\frac1n\TZ_{\Vtheta}^{\top}
 J\{G_0-G(\Vtheta)\}(\rho_0\Vy+X\Vbeta_{20})\right].
 \label{eq:network-discrepancy-new}
\end{align}
Proposition~\ref{Proposition: Valid Moment Condition} and the structural equation~\eqref{model:compact} imply
\begin{equation}\label{eq:moment-decomposition-new}
 m(\TVbeta,\Vtheta)
 =M(\Vtheta)(\TVbeta_0-\TVbeta)+d(\Vtheta).
\end{equation}
For any matrix $M$, let $M^+$ denote its Moore--Penrose inverse. Set
$\Pi_{\Vtheta}^{\perp}=I_l-M(\Vtheta)M(\Vtheta)^+$, the orthogonal projector onto the complement of the column space of $M(\Vtheta)$. The unrestricted population profile satisfies
\begin{equation}\label{eq:profile-identification-new}
 \inf_{\TVbeta\in\RR^p}\|m(\TVbeta,\Vtheta)\|^2
 =\|\Pi_{\Vtheta}^{\perp}\Vc_{\Vtheta}\|^2
 =\|\Pi_{\Vtheta}^{\perp}d(\Vtheta)\|^2.
\end{equation}

\begin{identificationassumption}[Global separation]\label{assumption: high level separation}
For every $\Vtheta\in\Theta\setminus\{\Vtheta_0\}$,
\[
 \Pi_{\Vtheta}^{\perp}d(\Vtheta)\ne\boldsymbol0.
\]
\end{identificationassumption}

This is a high-level condition requiring that no change in the mean coefficients can offset a false network in all NIV moments. It is a restriction for global uniqueness of the true parameters as the minimizer. When coefficients are restricted to a compact set $\mathcal B\subset\RR^p$ containing $\TVbeta_0$, the condition remains sufficient, although it can be stronger than necessary.

\begin{theorem}[Joint identification under global separation]\label{Theorem: General Identification}
Suppose Assumptions~\ref{assumption: conditional exogeneity} and \ref{assumption: Kernel Regularity} hold, $M(\Vtheta_0)$ has column rank $p$, and Assumption~\ref{assumption: high level separation} holds. Then $(\TVbeta_0,\Vtheta_0)$ is the unique zero of $m$ on $\RR^p\times\Theta$, and hence on $\mathcal B\times\Theta$.
\end{theorem}

The proof is given in Appendix~\ref{appendix: general identification}.

The following subsections give two alternative formulations sufficient for this result. The first establishes separation constructively from eigenvectors and eigenvalues of the network weighting matrices; while the second uses a span restriction on the moment blocks. The lettered assumptions describe these identification conditions separately from the numbered assumptions. They are not cumulative requirements for the two routes.

\subsection{Identification using common eigenvectors of the candidate network weighting matrices}

Let $\mathcal D$ denote the observed within-group distance matrices.
Under Assumption~\ref{assumption: Kernel Regularity}(ii),
$\sigma(\mathcal D)=\sigma(G_0)$, so the residualized instrument is
$R_Z=Z-\EE[Z\mid\mathcal D]$. Consider a unit vector $u$ that is an
eigenvector of $G(\Vtheta)$ for every $\Vtheta\in\Theta$. The vector
$u$ is common across candidates, although its eigenvalue
$\lambda(\Vtheta)$ may change with $\Vtheta$. Since $G(\Vtheta)$ is
symmetric, we have
\[
  u^\top G(\Vtheta)v=\lambda(\Vtheta)u^\top v.
\]
Thus, along a common eigenvector, the network spillover effect enters the structural
equation through a scalar eigenvalue. We use this observation to
identify the decay parameters.

\begin{identificationassumption}[Common eigenvectors and instrument relevance]
\label{assumption: extractable modes}
There exist $H\geq 4$ matrices $U_1,\ldots,U_H$, each with one or
more orthonormal columns, that are measurable functions of $\mathcal D$
and do not depend on $\Vtheta$. Almost surely, for every
$\Vtheta\in\Theta$,
\begin{equation}\label{eq:common-modes-new}
  U_s^\top U_t=0 \quad (s\neq t),\qquad
  JU_s=U_s,\qquad
  G(\Vtheta)U_s=\lambda_s(\Vtheta)U_s,
  \qquad s=1,\ldots,H.
\end{equation}
Write $P_s=U_sU_s^\top$ for the orthogonal projection onto the column space of $U_s$. The following conditions hold.
\begin{enumerate}[(i)]
\item Either $R_Z=\sum_{s=1}^H P_sR_Z$ almost surely, or
      $\sum_{s=1}^H P_s=J$.
\item The eigenvalue functions $\lambda_s:\Theta\to\RR$ are nonrandom,
      and $\lambda_s(\Vtheta)\neq\lambda_t(\Vtheta)$ for every
      $s\neq t$ and $\Vtheta\in\Theta$. The NIV stack in
      \eqref{equation: IV p choice} includes powers $0,\ldots,q$,
      with $q\geq H-1$.
\item For every $s$, the $k\times k$ matrix
      \[
        C_s\equiv
        \EE\!\left[\frac{1}{N}(U_s^\top R_Z)^\top(U_s^\top X)\right]
      \]
      is nonsingular. The corresponding cross-products with the
      projected outcome and disturbance are integrable.
\end{enumerate}
\end{identificationassumption}

The restriction $JU_s=U_s$ means that each column of $U_s$ sums to
zero within each group, so projection removes the group fixed effects.
Part~(i) requires the selected eigenvectors to span either the columns
of $R_Z$ or the entire space of within-group contrasts. Part~(iii) is a regular rank condition on the projected instruments and covariates.

The common eigenvectors and their eigenvalue functions can be checked
using the observed distances and the specified kernel. In particular,
for any column $u$ of $U_s$,
$\lambda_s(\Vtheta)=u^\top G(\Vtheta)u$ at each candidate $\Vtheta$.
Appendix~\ref{appendix: distance verification} gives a characterization
for the exponential kernel.

\begin{lemma}[NIV equations in the common eigenvectors]
\label{lemma: modal extraction}
Suppose Assumptions~\ref{assumption: conditional exogeneity},
\ref{assumption: Kernel Regularity}, and
\ref{assumption: extractable modes} hold. For $s=1,\ldots,H$, let
\[
  f_s\equiv C_s^{-1}
  \EE\!\left[\frac{1}{N}(U_s^\top R_Z)^\top(U_s^\top\Vy)\right]
  \in\RR^k.
\]
Then $m(\TVbeta,\Vtheta)=\boldsymbol 0$ if and only if
\begin{equation}\label{eq:modal-equations-new}
  \{1-\rho\lambda_s(\Vtheta)\}f_s
  =\Vbeta_1+\lambda_s(\Vtheta)\Vbeta_2,
  \qquad s=1,\ldots,H.
\end{equation}
In particular, whenever $1-\rho_0\lambda_s(\Vtheta_0)\neq 0$,
\begin{equation}\label{eq:modal-response-new}
  f_s=
  \frac{\Vbeta_{10}+\lambda_s(\Vtheta_0)\Vbeta_{20}}
       {1-\rho_0\lambda_s(\Vtheta_0)}.
\end{equation}
\end{lemma}

The vector $f_s$ is the population IV coefficient obtained from
the projected outcome $U_s^\top\Vy$, covariates $U_s^\top X$, and
instruments $U_s^\top R_Z$. The original NIV moments recover a
separate equation for each distinct eigenvalue because the $h$th
network power weights the $s$th equation by $\lambda_s(\Vtheta)^h$.
For $h=0,\ldots,H-1$, these weights form a nonsingular Vandermonde
matrix. Thus $H$ counts the distinct eigenvalues used in the pooled
moments across all groups.

To recover each group's decay parameter, we use four eigenvalues.
For four distinct real numbers, define their cross-ratio by
\begin{equation}\label{eq:cross-ratio-new}
  \mathcal C(t_1,t_2,t_3,t_4)
  \equiv
  \frac{(t_1-t_3)(t_2-t_4)}{(t_1-t_4)(t_2-t_3)}.
\end{equation}

\begin{identificationassumption}[Identification of decay from four eigenvalues]
\label{assumption: spectral shape}
For each group $r$, four of the matrices in
Assumption~\ref{assumption: extractable modes} have all their columns
supported on that group. In a fixed order, denote their eigenvalue
functions by $\lambda_{r1}(\theta_r),\ldots,\lambda_{r4}(\theta_r)$.
The function
\begin{equation}\label{eq:cross-ratio-shape-new}
  \chi_r(t)\equiv
  \mathcal C\bigl(\lambda_{r1}(t),\lambda_{r2}(t),
                  \lambda_{r3}(t),\lambda_{r4}(t)\bigr)
\end{equation}
is one-to-one on $\Theta_r$.
\end{identificationassumption}

Equation~\eqref{eq:modal-response-new} relates each coordinate of
$f_s$ to the corresponding eigenvalue through a ratio of two linear
functions. When this ratio is nonconstant, it preserves cross-ratios.
Consequently, taking the cross-ratio of four projected IV coefficients
in the same coordinate eliminates the structural coefficients.
Assumption~\ref{assumption: spectral shape} requires the remaining
quantity to distinguish every pair of decay parameters in that group.

\begin{theorem}[Joint identification from common eigenvectors and eigenvalues]
\label{Theorem: Spectral Identification}
Suppose Assumptions~\ref{assumption: conditional exogeneity},
\ref{assumption: Kernel Regularity},
\ref{assumption: extractable modes}, and
\ref{assumption: spectral shape} hold, and
\begin{equation}\label{eq:noncancellation-new}
  \Vc_0\equiv\Vbeta_{20}+\rho_0\Vbeta_{10}\neq\boldsymbol 0.
\end{equation}
Then $M(\Vtheta_0)$ has column rank $p=2k+1$ and
Assumption~\ref{assumption: high level separation} holds.
Consequently, $(\TVbeta_0,\Vtheta_0)$ is the unique zero of the
NIV moments on $\RR^p\times\Theta$. In particular,
let $f_{r\ell,j}$ be coordinate $j$ of the projected IV coefficient
corresponding to $\lambda_{r\ell}$. For any $j$ with $c_{0j}\neq 0$,
\begin{equation}\label{eq:theta-recovery-new}
  \theta_{0r}
  =\chi_r^{-1}\!\left(
     \mathcal C(f_{r1,j},f_{r2,j},f_{r3,j},f_{r4,j})
   \right),
  \qquad r=1,\ldots,\bar r.
\end{equation}
\end{theorem}

We provide the proofs in Appendix~\ref{Appendix: proof of identification using common eigenvectors}.

\subsection{Identification from the span restriction}

For general distance geometries, we have the following alternative sufficient condition for identification from the moment blocks. This is the second route.

\begin{assumption}[Separation from the false column space]\label{assumption: global separation universal}
Suppose $l > 2k+1$. For every $\Vtheta'\in\Theta\setminus\{\Vtheta_0\}$,
\[
B_{\Vtheta'}(\Vtheta_0)\Vgamma_0 \notin \operatorname{Span}\{A_{\Vtheta'},B_{\Vtheta'}(\Vtheta')\}.
\]
\end{assumption}

Proposition~\ref{Proposition: Valid Moment Condition} implies
\begin{equation}\label{eq:peer-decomposition-new}
 \Vc_{\Vtheta}=A_{\Vtheta}\Vbeta_{10}
              +B_{\Vtheta}(\Vtheta_0)\Vgamma_0.
\end{equation}
The columns of $[A_{\Vtheta},B_{\Vtheta}(\Vtheta)]$ are a permutation of those of $M(\Vtheta)$. Consequently, Assumption~\ref{assumption: global separation universal} is equivalent to Assumption~\ref{assumption: high level separation}: the true peer moment vector has a component outside the span generated by the direct covariates and the candidate-network peer regressors. This is an alternative formulation of separation.

\begin{theorem}[Joint identification under the span restriction]
    \label{Theorem: identification under span restriction}
    Under Assumptions \ref{assumption: conditional exogeneity}--\ref{assumption: global separation universal}, $\Vtheta_0$ and $\TVbeta_0$ are uniquely identified by the NIV moments.
\end{theorem}

We give the proof in Appendix~\ref{Appendix: proof of identification under span restriction}. Note that Assumptions~\ref{assumption: extractable modes} and \ref{assumption: spectral shape} are not required for this route. The kernel regulation in Assumption~\ref{assumption: Kernel Regularity}(iii) exclude a uniform rescaling of all network weights, but can not exclude cancellations after the outcome and covariate components have been projected onto the instruments. Appendix~\ref{appendix: gs sufficient} discusses the potential violation of Assumption~\ref{assumption: global separation universal}. Under some additional nondegeneracy conditions, possible false solutions form a null set in the candidate-parameter coordinates.

For either identification route, any positive definite population GMM weight $W$ yields a criterion $m(\TVbeta,\Vtheta)^{\top}Wm(\TVbeta,\Vtheta)$ with the unique zero $(\TVbeta_0,\Vtheta_0)$.

\subsection{Identification of the disturbance parameters}

Finally, once $(\TVbeta_0,\Vtheta_0)$ has been identified, define the structural residual
\[
\boldsymbol{u}_0\equiv \TVy-\TX\TVbeta_0=\TVep=(J+\delta_0JG(\Vtheta_0))\Vep.
\]
Under the conditional innovation covariance restriction in Assumption~\ref{assumption: covariance shape},
\[
\cov(\boldsymbol{u}_0\mid G_0)=\sigma_{\ep}^{2}\Sigma(\delta_0,\Vtheta_0),\qquad
\Sigma(\delta,\Vtheta_0)\equiv (J+\delta JG_0)(J+\delta JG_0)^{\top}.
\]
The covariance argument requires restrictions beyond network-conditional exclusion. We state them together with the shape restriction.
\begin{assumption}[Covariance Shape]\label{assumption: covariance shape}
The innovations satisfy
\[
 \cov(\Vep\mid G_0)=\sigma_{\ep}^2 I_N,
 \qquad 0<\sigma_{\ep}^2<\infty,
\]
and $\Sigma(\delta_0,\Vtheta_0)\ne\boldsymbol0$ almost surely. For every admissible $\delta\ne\delta_0$, there is no constant $a>0$ such that
\[
 \Sigma(\delta,\Vtheta_0)=a\,\Sigma(\delta_0,\Vtheta_0)
 \quad\text{almost surely}.
\]
\end{assumption}

\begin{theorem}[Identification of the disturbance parameters]\label{Theorem: disturbance identification}
If $(\TVbeta_0,\Vtheta_0)$ are identified and Assumption \ref{assumption: covariance shape} holds, then $\delta_0$ and $\sigma_{\ep}^{2}$ are identified from the covariance of $\boldsymbol{u}_0$.
\end{theorem}

The proof is straightforward, given in Appendix~\ref{Appendix: proof of disturbance identification}. Thus either Theorem~\ref{Theorem: Spectral Identification} or Theorem~\ref{Theorem: identification under span restriction}, together with Theorem~\ref{Theorem: disturbance identification}, identifies the mean coefficients, network decay parameters, and disturbance parameters in the model. The conditional innovation mean remains unrestricted, and the group fixed effects are not identified by these within equations.

\section{Estimation}\label{section: estimation}

We now describe our estimation and inference methods. Let
\[
 \eta=(\TVbeta^\top,\Vtheta^\top)^\top,\qquad
 d_\eta=2k+1+\bar r,\qquad
 u(\eta)=\TVy-\TX(\Vtheta)\TVbeta.
\]
We consider asymptotics as $N\to\infty$ and $n_r/N\to\pi_r>0$,
with $\bar r$, $k$, and $q$ fixed.

\subsection{Estimation and asymptotic results}

Let $\mathcal D$ retain its meaning as the observed within-group distance matrices, and define the network-information sigma-field
\[
 \mathcal F_G:=\sigma(\mathcal D)=\sigma(G_0)=\sigma(G(\Vtheta))
\]
for every admissible $\Vtheta$, under the information-equivalence conditions in
Section~\ref{section: identification}. Write $\mu_G=\EE[Z\mid\mathcal F_G]$
and $R_Z=Z-\mu_G$. The pooled population moment is
\begin{equation}\label{eq:joint-original-moment}
 m_N(\eta)=\EE\left[\frac1N\TZ_{\Vtheta}^\top u(\eta)\right].
\end{equation}
For the feasible moment $\hat m_N$ below and a symmetric positive definite
weight $W_N$ held fixed during optimization, define
\begin{equation}\label{eq:joint-estimator}
 \hat\eta\in\arg\min_{\eta\in\mathcal H}\hat Q_N(\eta),\qquad
 \hat Q_N(\eta)=\hat m_N(\eta)^\top W_N\hat m_N(\eta).
\end{equation}

For row $Z_{ri}$ and a vector or matrix response $R_r$, define
\[
 \begin{gathered}
 \mu_{ri}=\EE[Z_{ri}\mid\mathcal F_G],\qquad
 v_{rsi}(R_r,\theta_r)=e_i^\top\{G_r(\theta_r)^s\}^\top J_rR_r,\\
 a_{rsi}(R_r,\theta_r)=\EE[v_{rsi}(R_r,\theta_r)\mid\mathcal F_G].
 \end{gathered}
\]
For $s=0,\ldots,q$, the $s$th block of $\mathcal M_N^0(R,\Vtheta)$ is
\begin{equation}\label{eq:joint-oracle-operator}
 [\mathcal M_N^0(R,\Vtheta)]_s
 =\frac1N\sum_{r,i}(Z_{ri}-\mu_{ri})^\top
 \{v_{rsi}(R_r,\theta_r)-a_{rsi}(R_r,\theta_r)\}.
\end{equation}
Stack the blocks in order $s=0,\ldots,q$. For $R_x(\Vtheta)=(G(\Vtheta)\Vy,X,G(\Vtheta)X)$, let
\[
 m_N^0(\eta)=\mathcal M_N^0(\Vy,\Vtheta)
             -\mathcal M_N^0\{R_x(\Vtheta),\Vtheta\}\TVbeta.
\]
Conditional centering gives $\EE[m_N^0(\eta)]=m_N(\eta)$ at every $\eta$.

Replace the conditional means in \eqref{eq:joint-oracle-operator} by
out-of-fold predictions to obtain $\widehat{\mathcal M}_N$, and set
\begin{equation}\label{eq:joint-feasible-moment}
 \hat m_N(\eta)=\widehat{\mathcal M}_N(\Vy,\Vtheta)
       -\widehat{\mathcal M}_N\{R_x(\Vtheta),\Vtheta\}\TVbeta.
\end{equation}
For the regressions within group $r$, one choice is to smooth on an
observed index $T_{ri}$ of fixed dimension. This requires
$\mu_{ri}=\mu_r(T_{ri})$ and
$a_{rsi}(R_r,\theta_r)=a_{rs}(T_{ri};R_r,\theta_r)$
and the smoothness and support conditions for local-constant or
local-polynomial regression. We use the same folds and linear
smoothing weights for $\Vy$ and the columns of $R_x(\Vtheta)$.

For
\[
 w_{rsi}(\eta)=v_{rsi}(\Vy_r,\theta_r)
       -v_{rsi}\{R_{x,r}(\theta_r),\theta_r\}\TVbeta,\qquad
 a_{rsi}(\eta)=\EE[w_{rsi}(\eta)\mid\mathcal F_G],
\]
the summand in the $s$th block is
$(Z_{ri}-\mu_{ri})^\top\{w_{rsi}(\eta)-a_{rsi}(\eta)\}$.

Its expectation has zero directional derivative with respect to either
conditional mean at the population values
\citep{ChernozhukovEtAl2018JointNote}.
At $\eta_0$, put $\Delta\mu_{ri}=\hat\mu_{ri}-\mu_{ri}$ and
$\Delta a_{rsi}=\hat a_{rsi}(\eta_0)-a_{rsi}(\eta_0)$.
The $s$th block of $\hat m_N(\eta_0)-m_N^0(\eta_0)$ equals
\begin{equation}\label{eq:joint-remainder}
 -\frac1N\sum_{r,i}\Delta\mu_{ri}^\top(w_{rsi}-a_{rsi})
 -\frac1N\sum_{r,i}(Z_{ri}-\mu_{ri})^\top\Delta a_{rsi}
 +\frac1N\sum_{r,i}\Delta\mu_{ri}^\top\Delta a_{rsi}.
\end{equation}
If each of the first two sums is $o_p(N^{-1/2})$ and
\begin{equation}\label{eq:joint-nuisance-rates}
 \begin{gathered}
 r_{\mu,N}^2=\frac1N\sum_{r,i}\|\Delta\mu_{ri}\|^2=o_p(1),\qquad
 r_{a,N}^2=\max_{0\le s\le q}\frac1N\sum_{r,i}|\Delta a_{rsi}|^2=o_p(1),\\
 \sqrt N r_{\mu,N}r_{a,N}=o_p(1),
 \end{gathered}
\end{equation}
then $\sqrt N\{\hat m_N(\eta_0)-m_N^0(\eta_0)\}=o_p(1)$ by
Cauchy--Schwarz.

\begin{assumption}[Regularity conditions for consistency]\label{ass:joint-consistency}
\leavevmode
\begin{enumerate}
 \item[(i)] $\mathcal H$ is compact and contains $\eta_0$.
 The map $m_N$ converges uniformly on $\mathcal H$ to a continuous $m$,
 and $m(\eta)=0$ if and only if $\eta=\eta_0$.
 \item[(ii)] $\sup_{\eta\in\mathcal H}\|\hat m_N(\eta)-m_N(\eta)\|=o_p(1)$
 and $W_N\to_p W$, where $W$ is symmetric positive definite.
 \item[(iii)] $\hat Q_N(\hat\eta)\le\inf_{\eta\in\mathcal H}\hat Q_N(\eta)+o_p(1)$.
\end{enumerate}
\end{assumption}

An oracle uniform law, uniform empirical $L^2$ consistency of the
nuisance estimates, and bounded empirical second moments suffice for
the moment convergence in part (ii). See
\citet[Theorem~2]{JenishPrucha2009JointNote} for a spatial uniform law.

\begin{assumption}[Regularity conditions for asymptotic normality]\label{ass:joint-regular}
\leavevmode
\begin{enumerate}
 \item[(i)] $\eta_0$ is interior to $\mathcal H$. On a fixed neighborhood
 $\mathcal U$ of $\eta_0$, $\hat m_N$ is continuously differentiable and
 \[
  \sup_{\eta\in\mathcal U}
       \|\partial_\eta\hat m_N(\eta)-D(\eta)\|=o_p(1),
 \]
 where $D(\eta)$ is continuous and $D=D(\eta_0)$ has rank $d_\eta$.
 \item[(ii)] $\sqrt N m_N^0(\eta_0)\Rightarrow\mathcal N(0,\Omega)$,
 with $\Omega$ finite and positive definite.
 Each of the first two sums in \eqref{eq:joint-remainder} is
 $o_p(N^{-1/2})$, and \eqref{eq:joint-nuisance-rates} holds.
 \item[(iii)] The approximate first-order condition satisfies
 \[
  \|\partial_\eta\hat m_N(\hat\eta)^\top W_N\hat m_N(\hat\eta)\|
       =o_p(N^{-1/2}).
 \]
\end{enumerate}
\end{assumption}

Full column rank in part (i) requires at least $d_\eta$ pooled moments
\citep{Lee2007JointNote}:
\begin{equation}\label{eq:joint-count}
 (q+1)k\ge 2k+1+\bar r.
\end{equation}

As in \citet{Lee2007JointNote}, moment restrictions on $X$, $Z$, and
$\Vep$ and uniformly bounded absolute row and column sums of $G$, its
derivatives, and the outcome filter give the score and derivative
envelopes. Since $\bar r$ is fixed, the score CLT is a within-group
limit. One may apply
\citet[Theorem~3.2]{KojevnikovMarmerSong2021JointNote} when the score
array satisfies their dependence-decay and neighborhood-growth
conditions.

The first two terms in \eqref{eq:joint-remainder} require control of
dependence between the training and held-out samples.

\begin{theorem}[NIV-GMM]\label{thm:joint-gmm}
Under Assumption~\ref{ass:joint-consistency}, $\hat\eta\to_p\eta_0$.
If Assumption~\ref{ass:joint-regular} also holds, then
\begin{align}
 \sqrt N(\hat\eta-\eta_0)
   &=-A\sqrt N m_N^0(\eta_0)+o_p(1),
 & A&=(D^\top WD)^{-1}D^\top W,\label{eq:joint-if}\\
 \sqrt N(\hat\eta-\eta_0)&\Rightarrow\mathcal N(0,V),
 & V&=A\Omega A^\top.\label{eq:joint-sandwich}
\end{align}
For $W=\Omega^{-1}$, $V=(D^\top\Omega^{-1}D)^{-1}$.
\end{theorem}

\subsection{Feasible inference and hypothesis testing}\label{sec:joint-inference}

Let $l=(q+1)k$ and let $\psi_i$ denote the $l$-vector oracle score at $\eta_0$, so that
$m_N^0(\eta_0)=N^{-1}\sum_i\psi_i$ and
$\EE[\psi_i\mid\mathcal F_G]=0$.
\[
 \begin{aligned}
 \psi_{\iota(r,i)}
 &:=\begin{bmatrix}
 (Z_{ri}-\mu_{ri})^\top\{w_{r0i}(\eta_0)-a_{r0i}(\eta_0)\}\\
 \vdots\\
 (Z_{ri}-\mu_{ri})^\top\{w_{rqi}(\eta_0)-a_{rqi}(\eta_0)\}
 \end{bmatrix},\\
 \hat\psi_{\iota(r,i)}
 &:=\begin{bmatrix}
 (Z_{ri}-\hat\mu_{ri})^\top\{w_{r0i}(\hat\eta)-\hat a_{r0i}(\hat\eta)\}\\
 \vdots\\
 (Z_{ri}-\hat\mu_{ri})^\top\{w_{rqi}(\hat\eta)-\hat a_{rqi}(\hat\eta)\}
 \end{bmatrix},\\
 \iota(r,i)&:=\sum_{t<r}n_t+i,\qquad i=1,\ldots,n_r.
 \end{aligned}
\]
When first-step estimation contributes at first order, an alternative to Assumption~\ref{ass:joint-regular}(ii) is the existence of $l$-vector scores $\psi_i^{\mathrm{full}}$ satisfying
\begin{equation}\label{eq:joint-full-score}
 \sqrt N\hat m_N(\eta_0)
 =\frac1{\sqrt N}\sum_i\psi_i^{\mathrm{full}}+o_p(1),
 \qquad
 \frac1{\sqrt N}\sum_i\psi_i^{\mathrm{full}}
 \Rightarrow\mathcal N(0,\Omega_{\mathrm{full}}),
\end{equation}
where $\Omega_{\mathrm{full}}$ is finite and positive definite. Under Assumption~\ref{ass:joint-consistency} and parts (i) and (iii) of Assumption~\ref{ass:joint-regular}, the same Taylor expansion then gives
\[
 \sqrt N(\hat\eta-\eta_0)
 =-\frac A{\sqrt N}\sum_i\psi_i^{\mathrm{full}}+o_p(1)
 \Rightarrow\mathcal N(0,A\Omega_{\mathrm{full}}A^\top).
\]
In this case, use $\psi_i=\psi_i^{\mathrm{full}}$, $\Omega=\Omega_{\mathrm{full}}$, and $V=A\Omega_{\mathrm{full}}A^\top$ below, and additionally require $\EE[\psi_i^{\mathrm{full}}\mid\mathcal F_G]=0$ for the conditional covariance construction.
Let $\hat\psi_i$ be a feasible estimate of the selected score, including estimated first-step influence contributions in the full-score case. For symmetric,
$\mathcal F_G$-measurable weights $\omega_{N,ij}$, define
\begin{equation}\label{eq:joint-hac}
 \hat\Omega_N=\frac1N\sum_{i,j}\omega_{N,ij}\hat\psi_i\hat\psi_j^\top.
\end{equation}
A common choice is $\omega_{N,ij}=\kappa(\ell_{ij}/b_N)$, where
$\ell_{ij}$ is an observed distance along which score dependence,
including dependence induced by network errors, decays.

For consistency, assume the following bounds, with covariances
conditional on $\mathcal F_G$,
\begin{align}
 &\left\|\frac1N\sum_{i,j}(1-\omega_{N,ij})
                  \EE[\psi_i\psi_j^\top\mid\mathcal F_G]\right\|=o_p(1),
                  \label{eq:joint-hac-bias}\\
 &\frac1{N^2}\sum_{i,j,k,t}\omega_{N,ij}\omega_{N,kt}
  \operatorname{Cov}(\psi_{i,a}\psi_{j,b},\psi_{k,a}\psi_{t,b}
                                      \mid\mathcal F_G)=o_p(1)
  \quad\text{for every }a,b,\label{eq:joint-hac-variance}\\
 &\frac1N\sum_{i,j}\EE[\psi_i\psi_j^\top\mid\mathcal F_G]\to_p\Omega,
 \qquad \frac1N\sum_i\|\psi_i\|^2=O_p(1),\label{eq:joint-hac-limit}\\
 &e_N^2=\frac1N\sum_i\|\hat\psi_i-\psi_i\|^2=o_p(1),\qquad
 B_Ne_N=o_p(1),\qquad B_N=\max_i\sum_j|\omega_{N,ij}|.
 \label{eq:joint-hac-plugin}
\end{align}
\citet{KelejianPrucha2007JointNote} and
\citet[Proposition~4.1]{KojevnikovMarmerSong2021JointNote} relate bandwidth
growth to dependence and neighborhood size.
These bounds give $\hat\Omega_N\to_p\Omega$. Under
Assumptions~\ref{ass:joint-consistency}--\ref{ass:joint-regular}, or their full-score replacement stated above, compute
$\hat D_N=\partial_\eta\hat m_N(\hat\eta)$ by differentiating both the
instruments and the residuals, and define
\begin{equation}\label{eq:joint-feasible-covariance}
 \hat A_N=(\hat D_N^\top W_N\hat D_N)^{-1}\hat D_N^\top W_N,\qquad
 \hat V_N=\hat A_N\hat\Omega_N\hat A_N^\top.
\end{equation}
Then $\hat V_N\to_p V$.

We may set $\omega_{N,ij}$ to zero between groups whenever the groups
are independent or their aggregate cross-covariance is asymptotically
negligible.

Under Theorem~\ref{thm:joint-gmm} or the full-score extension, and
\eqref{eq:joint-hac-bias}--\eqref{eq:joint-hac-plugin}, Wald inference
follows by the delta method.
Let $h:\mathcal H\to\RR^{s_h}$ be continuously differentiable near
$\eta_0$, with $h(\eta_0)=0$ and
$H_0VH_0^\top\succ0$, where $H_0=\partial_\eta h(\eta_0)$.
For $\hat H=\partial_\eta h(\hat\eta)$,
\begin{equation}\label{eq:joint-wald}
 N h(\hat\eta)^\top(\hat H\hat V_N\hat H^\top)^{-1}h(\hat\eta)
       \Rightarrow\chi^2_{s_h}.
\end{equation}
A $(1-\alpha)$ confidence interval for $\eta_{0j}$
is $\hat\eta_j\pm z_{1-\alpha/2}\sqrt{[\hat V_N]_{jj}/N}$.
For $W=\Omega^{-1}$, partitioning $H=D^\top\Omega^{-1}D$ by
$(\TVbeta,\Vtheta)$ gives the coefficient covariance
\[
 V_{\beta\beta}
  =(H_{\beta\beta}-H_{\beta\theta}H_{\theta\theta}^{-1}
                                      H_{\theta\beta})^{-1}.
\]

If $W_N(\eta)$ converges uniformly to a continuous positive definite
$W(\eta)$ and its first derivatives are locally $O_p(1)$, the same
expansion holds at a consistent approximate stationary point, using
$W=W(\eta_0)$.

For a Gaussian--logistic kernel mixture, consider weight
$a\in[0,1]$ on the logistic kernel. At $a=0$ and $a=1$, the mixture
weight lies on the boundary of its parameter space, outside the scope
of the interior Wald result. One may fix $a$ at each value in a finite grid
$\mathcal A$ and use the smallest interval containing the resulting
coefficient intervals. If
$a_0\in\mathcal A$, all grid intervals can be computed with probability
tending to one, and the interval at $a_0$ has nominal asymptotic
coverage, the envelope has at least nominal asymptotic coverage
(see Appendix~\ref{sec:joint-proofs}).

For the test for mean network spillovers, consider\label{sec:restricted-niv}
\begin{equation}\label{eq:restricted-null}
 H_0:\quad \rho_0=0,\qquad\Vbeta_{20}=\mathbf0.
\end{equation}
Under this null $\rho_0\Vy+X\Vbeta_{20}=0$, so the decay columns of the
population Jacobian vanish by \eqref{eq:joint-theta-derivative}. After
removing group effects, the model under the null is
\[
 J(\Vy-X\Vbeta_{10})=J(I_N+\delta_0G_0)\Vep,
\]
with $\delta_0$ unrestricted.

We therefore fix a reference value $\bar{\Vtheta}$ before examining the
outcomes or instruments and construct $G(\bar{\Vtheta})$ from the observed geometry.
Under conditional exclusion and information equivalence, the reference
moments are valid under the null even if $\bar{\Vtheta}\ne\Vtheta_0$.
Define
\[
 \hat c=\widehat{\mathcal M}_N(\Vy,\bar{\Vtheta}),\qquad
 \hat M=\widehat{\mathcal M}_N(X,\bar{\Vtheta}),\qquad
 \hat g(b)=\hat c-\hat M b.
\]
The subscript $0$ denotes the own-instrument block, and $+$ denotes the
other $qk$ rows.
The restricted estimator
$\hat b=\hat M_0^{-1}\hat c_0$ is just-identified IV on the residualized
within-group variables.

Prespecify the $k\times qk$ contrast
$C=(2^{-1}I_k,\ldots,2^{-q}I_k)$, with $q\ge1$. Put $d=k$ and
\begin{align}
 \hat L&=(-\hat M_+\hat M_0^{-1},I_{qk}),&
 \hat z&=C\hat g_+(\hat b),\nonumber\\
 \hat V_C&=C\hat L\hat\Omega\hat L^\top C^\top,&
 J_C&=N\hat z^\top\hat V_C^{-1}\hat z.
 \label{eq:restricted-stat}
\end{align}
Here $\hat\Omega$ estimates the long-run covariance of
$\sqrt N\hat g(\Vbeta_{10})$, including required first-step terms.
This is a test of fixed overidentifying restrictions \citep{Hansen1982}.

\begin{assumption}[Inference for the test for mean network spillovers]\label{ass:restricted-niv}
Under \eqref{eq:restricted-null}, maintain conditional exclusion,
information equivalence, and integrability of the displayed moments.
At the prespecified reference:
\begin{enumerate}
 \item[(i)] $\hat M\to_p M$, with deterministic $M$ and nonsingular $M_0$.
 \item[(ii)] $\sqrt N\hat g(\Vbeta_{10})\Rightarrow\mathcal N(0,\Omega)$.
 \item[(iii)] $\hat\Omega\to_p\Omega$ and
 $V_C=CL\Omega L^\top C^\top\succ0$, where
 $L=(-M_+M_0^{-1},I_{qk})$.
\end{enumerate}
\end{assumption}

For the CLT and covariance in parts (ii)--(iii), use the score conditions and
\eqref{eq:joint-remainder}--\eqref{eq:joint-nuisance-rates} at the
reference parameter, or the corresponding conditions for an
adjusted score.

\begin{theorem}[Test for mean network spillovers]\label{thm:restricted-niv}
Under Assumption~\ref{ass:restricted-niv},
$J_C\Rightarrow\chi^2_d$. The rule
$J_C>\chi^2_{d,1-\alpha}$ has limiting rejection probability $\alpha$
at each null law satisfying the assumption.
The result also holds if parts (i)--(iii) hold conditionally for almost
every realization of a common design environment.

Under alternatives, suppose the matrix and covariance limits in
Assumption~\ref{ass:restricted-niv} hold along a sequence of alternatives,
and, for a deterministic sequence $b_N$,
$\sqrt N\hat g(b_N)\Rightarrow\mathcal N(h,\Omega)$. Then
\[
 J_C\Rightarrow\chi^2_d(\lambda),\qquad
 \lambda=(CLh)^\top V_C^{-1}(CLh).
\]
Under a fixed alternative with $\hat c\to_p c$, $\hat M\to_p M$,
$M_0$ nonsingular, and $\hat V_C\to_p V_A\succ0$, let
$z_A=C(c_+-M_+M_0^{-1}c_0)$. Then
$J_C/N\to_p z_A^\top V_A^{-1}z_A$.
Consequently, rejection probability tends to one whenever $z_A\ne0$.
\end{theorem}

With zero group effects, a stable $\rho_0\ne0$, and
$\Vbeta_{20}=-\rho_0\Vbeta_{10}$, the model reduces to
\[
 \Vy=X\Vbeta_{10}
       +(I_N-\rho_0G_0)^{-1}(I_N+\delta_0G_0)\Vep.
\]
The population moments are zero at $\Vbeta_{10}$ under this alternative.
The exclusion and outcome specification remain
maintained when interpreting rejection of \eqref{eq:restricted-null}.

Proofs for this section are given in Appendix~\ref{sec:joint-proofs}.

\section{Simulation}\label{section: simulation}
We fit a Gaussian kernel and a mixture of Gaussian and logistic kernels
to each simulated sample. We use four group sizes,
$n_r=100,200,300,500$, with ten groups at each size. Within each
group we draw five covariates and arrange the units in cells of
six or seven. The first six units in each cell form three pairs.
The true network uses Gaussian or logistic
weights for two units at squared distance $d$, using the convention of Section~\ref{section: model setup},
\[
 K_{\theta_r}^{\mathrm G}(d)=e^{-\theta_r d},
 \qquad K_{\theta_r}^{\mathrm L}(d)=\frac{2e^{-\theta_r d}}{1+e^{-\theta_r d}}.
\]
The two designs use the same decay $c_r=0.82+0.38(r-1)/9$ in group $r$.
The remaining parameters of the outcome equation are
\[
 \begin{gathered}
 (\rho_0,\delta_0)=(0.28,0.15),\\
 \boldsymbol{\beta}_{10}=(1,0.75,1.20,0.85,1.05)^\top,\qquad
 \boldsymbol{\beta}_{20}=(2.20,1.60,1.80,1.45,1.95)^\top.
 \end{gathered}
\]

In estimation the group decays take the constrained form $\theta_r=c_r\vartheta$,
with known relative decays $c_r$ and one unknown common multiplier $\vartheta$.
Thus the Gaussian fit has 12 parameters and the free-mixture fit has 13,
rather than estimating all ten decays separately.
Both fits use $K_{\theta_r,a}=(1-a)K_{\theta_r}^{\mathrm G}
+aK_{\theta_r}^{\mathrm L}$. We estimate $\rho$, $\boldsymbol{\beta}_1$,
$\boldsymbol{\beta}_2$, and $\vartheta$ with $a=0$ for the Gaussian fit.
For the mixture, $a\in[0,1]$ is estimated too. For each of the five
instruments we use the centered instrument and its transforms by
$G_r^s$ for $s=1,\ldots,4$. We pool the resulting 25 moments and
use the Gaussian reference weight for both fits. For a held-out cell, we predict the
instrument and transformed residual means from the training cells
by nonlinear least squares and local linear regression, respectively.

Write $[L_j(a),U_j(a)]$ for the nominal 95\% one-step interval for
coefficient $j$ at fixed $a$, with cell-level standard errors and
both first-step corrections. We report the interval at $a=0$ for
the Gaussian fit and
\[
 [\min_{a\in\mathcal A} L_j(a),\ \max_{a\in\mathcal A} U_j(a)],
 \qquad \mathcal A=\{0,1/8,\ldots,1\},
\]
for the mixture. At each $n_r$, we use the same 200 replications
for point estimates and intervals.

\begin{table}[H]
\centering
\begin{threeparttable}
\caption{Bias, standard deviation, and RMSE of GMM-NIV}\label{tab:point}
\small\setlength{\tabcolsep}{5pt}
\begin{tabular*}{0.93\linewidth}{@{\extracolsep{\fill}}r l rrr rrr@{}}
\toprule
& & \multicolumn{3}{c}{Direct effects ($\boldsymbol{\beta}_1$)}
    & \multicolumn{3}{c}{Spillover effects ($\boldsymbol{\beta}_2$)} \\
\cmidrule(lr){3-5}\cmidrule(lr){6-8}
$n_r$ & Fit & Bias & SD & RMSE & Bias & SD & RMSE \\
\midrule
\multicolumn{8}{@{}l}{\textit{Panel A. Gaussian truth}}\\[2pt]
100 & Gaussian & -0.004 & 0.087 & 0.087 & 0.051 & 0.370 & 0.377 \\
 & Mixture & -0.006 & 0.088 & 0.088 & 0.027 & 0.361 & 0.365 \\
\addlinespace[2pt]
200 & Gaussian & -0.002 & 0.047 & 0.047 & 0.006 & 0.201 & 0.201 \\
 & Mixture & -0.003 & 0.048 & 0.050 & -0.014 & 0.204 & 0.204 \\
\addlinespace[2pt]
300 & Gaussian & 0.001 & 0.036 & 0.036 & 0.006 & 0.149 & 0.149 \\
 & Mixture & 0.000 & 0.039 & 0.040 & -0.012 & 0.151 & 0.151 \\
\addlinespace[2pt]
500 & Gaussian & 0.000 & 0.025 & 0.025 & -0.003 & 0.111 & 0.111 \\
 & Mixture & -0.001 & 0.026 & 0.026 & -0.011 & 0.114 & 0.115 \\
\addlinespace[6pt]
\multicolumn{8}{@{}l}{\textit{Panel B. Logistic truth}}\\[2pt]
100 & Gaussian & 0.041 & 0.184 & 0.290 & 0.375 & 0.424 & 0.576 \\
 & Mixture & 0.027 & 0.209 & 0.263 & 0.267 & 0.419 & 0.505 \\
\addlinespace[2pt]
200 & Gaussian & 0.053 & 0.119 & 0.261 & 0.347 & 0.251 & 0.431 \\
 & Mixture & 0.022 & 0.142 & 0.179 & 0.172 & 0.275 & 0.325 \\
\addlinespace[2pt]
300 & Gaussian & 0.059 & 0.092 & 0.250 & 0.332 & 0.193 & 0.386 \\
 & Mixture & 0.017 & 0.105 & 0.124 & 0.099 & 0.215 & 0.237 \\
\addlinespace[2pt]
500 & Gaussian & 0.057 & 0.067 & 0.243 & 0.326 & 0.147 & 0.360 \\
 & Mixture & 0.009 & 0.071 & 0.084 & 0.071 & 0.158 & 0.173 \\
\bottomrule
\end{tabular*}
\begin{tablenotes}[flushleft]\footnotesize
\item \textit{Notes.} $n_r$ is the number of units per group. Bias is the
average of the five signed coefficient biases within each block.
SD is the root mean square of the five standard deviations across
replications. RMSE pools squared errors over the five coefficients and
the available replications.
\end{tablenotes}
\end{threeparttable}
\end{table}

The two fits have similar RMSEs under Gaussian truth
(Table~\ref{tab:point}). The RMSEs of both direct effects and spillovers
fall as $n_r$ increases. Under logistic truth the mixture has lower
RMSEs at every group size. At $n_r=500$ the spillover RMSE is 0.360
for the Gaussian fit and 0.173 for the mixture. Gaussian fit has an increasing bias for direct effects and a persistent one for the spillovers, while mixture fit has decreasing bias and sd. 

\begin{table}[H]
\centering
\begin{threeparttable}
\caption{Coverage and mean width of nominal 95\% intervals}\label{tab:intervals}
\footnotesize\setlength{\tabcolsep}{4pt}
\begin{tabular*}{\linewidth}{@{\extracolsep{\fill}}r rrrr rrrr@{}}
\toprule
&\multicolumn{4}{c}{Gaussian fit}&\multicolumn{4}{c}{Mixture fit}\\
\cmidrule(lr){2-5}\cmidrule(lr){6-9}
&\multicolumn{2}{c}{Direct ($\boldsymbol{\beta}_1$)}
&\multicolumn{2}{c}{Spillover ($\boldsymbol{\beta}_2$)}
&\multicolumn{2}{c}{Direct ($\boldsymbol{\beta}_1$)}
&\multicolumn{2}{c}{Spillover ($\boldsymbol{\beta}_2$)}\\
\cmidrule(lr){2-3}\cmidrule(lr){4-5}\cmidrule(lr){6-7}\cmidrule(lr){8-9}
$n_r$ & Coverage & Width & Coverage & Width & Coverage & Width & Coverage & Width\\
\midrule
\multicolumn{9}{@{}l}{\textit{Panel A. Gaussian truth}}\\[2pt]
100 & 83.0 & 0.25 & 90.4 & 1.04 & 89.9 & 0.31 & 94.4 & 1.22 \\
200 & 92.2 & 0.16 & 95.0 & 0.66 & 95.2 & 0.22 & 96.7 & 0.83 \\
300 & 93.0 & 0.13 & 95.6 & 0.51 & 97.0 & 0.18 & 97.6 & 0.69 \\
500 & 94.6 & 0.10 & 94.5 & 0.38 & 97.5 & 0.15 & 96.7 & 0.56 \\
\addlinespace[5pt]
\multicolumn{9}{@{}l}{\textit{Panel B. Logistic truth}}\\[2pt]
100 & 38.8 & 0.42 & 66.8 & 1.20 & 91.6 & 0.64 & 95.2 & 1.50 \\
200 & 15.5 & 0.28 & 43.6 & 0.76 & 93.9 & 0.49 & 97.4 & 1.05 \\
300 & 5.2 & 0.23 & 32.1 & 0.59 & 94.8 & 0.43 & 97.3 & 0.87 \\
500 & 0.3 & 0.18 & 21.5 & 0.44 & 95.4 & 0.37 & 98.2 & 0.73 \\

\bottomrule
\end{tabular*}
\begin{tablenotes}[flushleft]\footnotesize
\item \textit{Notes.} Coverage is in percent. For each block we average
coverage and length over its five coefficients and the available intervals. For the
mixture we use replications with all nine intervals.
\end{tablenotes}
\end{threeparttable}
\end{table}

The intervals get shorter as $n_r$ grows (Table~\ref{tab:intervals}).
Under Gaussian truth the Gaussian direct-effect intervals cover
83.0\% at $n_r=100$ and 94.6\% at $n_r=500$. At $n_r=500$ under
logistic truth, spillover coverage is 21.5\% for the Gaussian fit
and 98.2\% for the mixture. The corresponding mean widths are 0.44
and 0.73.

For the Section~5 test for mean network spillovers, we estimate the direct effects by IV under
$\rho=0$ and $\boldsymbol{\beta}_2=\mathbf 0$. At the Gaussian reference
$\vartheta=1$, we combine the residual moments for each instrument
with weights $2^{-s}$ for $s=1,\ldots,4$ and use the 0.95 quantile of
$\chi^2_5$. Size and power are the rejection frequencies in 2,000 null
and 200 alternative replications at each group size, with $\delta_0=0.15$
in both. At size 0.05, the larger null batch gives a Monte Carlo
standard error of about 0.005.

In Table~\ref{tab:mean_null}, size rises with $n_r$ and remains below
0.05 under both kernels. The test rejects in all 200 replications
of the specified alternative at every group size.

\begin{table}[H]
\centering
\begin{threeparttable}
\caption{Rejection probabilities for the test for mean network spillovers}\label{tab:mean_null}
\small
\begin{tabular*}{0.88\linewidth}{@{\extracolsep{\fill}}r rr rr@{}}
\toprule
&\multicolumn{2}{c}{Gaussian truth}&\multicolumn{2}{c}{Logistic truth}\\
\cmidrule(lr){2-3}\cmidrule(lr){4-5}
$n_r$ & Size & Power & Size & Power\\
\midrule
100 & 0.0050 & 1.0000 & 0.0050 & 1.0000 \\
200 & 0.0115 & 1.0000 & 0.0110 & 1.0000 \\
300 & 0.0150 & 1.0000 & 0.0140 & 1.0000 \\
500 & 0.0275 & 1.0000 & 0.0270 & 1.0000 \\

\bottomrule
\end{tabular*}
\begin{tablenotes}[flushleft]\footnotesize
\item \textit{Notes.} The nominal level is 0.05. The alternative uses
the values of $\rho_0$ and $\boldsymbol{\beta}_{20}$ given above.
\end{tablenotes}
\end{threeparttable}
\end{table}

The data-generating process and implementation details are given in
Appendix~\ref{app:peer_cell}.

\section{Conclusion}\label{section: conclusion}

This paper develops network-adaptive instrumental variables for spatial autoregressive models with endogenous covariates and unknown network weights. Under the stated exclusion and information-equivalence assumptions, NIV yields valid moments without observing the true network weights. Joint identification of the structural mean coefficients and the network parameters follows a global separation condition or two alternative sufficient routes, while separate covariance restrictions identify the disturbance parameters. Under additional regularity conditions, the NIV-GMM estimator is consistent and asymptotically normal. We provide feasible covariance estimation, confidence intervals, and a test for mean network spillovers based on fixed moment restrictions. Monte Carlo simulations examine estimation and inference under Gaussian and logistic network kernels, including the finite-sample behavior of a kernel-mixture procedure.

\newpage

\appendix\label{appendix}
\section{Proofs}\label{appendix: proofs}
\label{appendix: general identification}
\label{Appendix: proof of NIV validity}
\label{Appendix: proof of identification using common eigenvectors}
\label{appendix: distance verification}
\label{Appendix: proof of identification under span restriction}
\label{Appendix: proof of disturbance identification}
\label{sec:joint-proofs}


\begin{proof}[Proof of Theorem~\ref{Theorem: General Identification}]
Proposition~\ref{Proposition: Valid Moment Condition} implies
$m(\TVbeta_0,\Vtheta_0)=\boldsymbol0$. Consider any
$(\TVbeta,\Vtheta)\in\RR^p\times\Theta$ satisfying
$m(\TVbeta,\Vtheta)=\boldsymbol0$. The structural equation gives
\[
\TVy-\TX(\Vtheta)\TVbeta
=
\TX(\Vtheta)(\TVbeta_0-\TVbeta)
+
J\{G_0-G(\Vtheta)\}
(\rho_0\Vy+X\Vbeta_{20})
+
\TVep.
\]
Premultiplying by $n^{-1}\TZ_{\Vtheta}^{\top}$, taking
expectations, and applying
Proposition~\ref{Proposition: Valid Moment Condition} yield
\[
\boldsymbol0
=
m(\TVbeta,\Vtheta)
=
M(\Vtheta)(\TVbeta_0-\TVbeta)+d(\Vtheta).
\]
Since $\Pi_{\Vtheta}^{\perp}M(\Vtheta)=0$, premultiplication by
$\Pi_{\Vtheta}^{\perp}$ gives
\[
\Pi_{\Vtheta}^{\perp}d(\Vtheta)=\boldsymbol0.
\]
Assumption~\ref{assumption: high level separation} therefore
implies $\Vtheta=\Vtheta_0$. At this value,
$d(\Vtheta_0)=\boldsymbol0$, so
\[
M(\Vtheta_0)(\TVbeta_0-\TVbeta)=\boldsymbol0.
\]
The full column rank of $M(\Vtheta_0)$ implies
$\TVbeta=\TVbeta_0$. Thus $(\TVbeta_0,\Vtheta_0)$ is the unique
zero of $m$ on $\RR^p\times\Theta$, and consequently on
$\mathcal B\times\Theta$.
\end{proof}


\begin{proof}[Proof of Proposition~\ref{Proposition: Valid Moment Condition}]
Fix $\Vtheta,\Vtheta'\in\Theta$ and a finite nonnegative integer
$q$. By Assumption~\ref{assumption: Kernel Regularity}(ii), for each
group $r$ there exists a Borel measurable function $\varphi_r$
such that
\[
K_{\theta_r}(d)=\varphi_r\{K_{\theta_r'}(d)\}
\]
on the support of the within-group off-diagonal distances.
Then we have
\[
g_{r,ij}(\theta_r)
=
a_{r,n}\varphi_r
\left(\frac{g_{r,ij}(\theta_r')}{a_{r,n}}\right)
\qquad\text{almost surely}.
\]
Since the group partition and scaling factors are fixed, and
the diagonal and cross-group entries are zero, $G(\Vtheta)$ is
a Borel measurable function of $G(\Vtheta')$. Thus
\[
\sigma\{G(\Vtheta)\}\subseteq\sigma\{G(\Vtheta')\}.
\]
Interchanging $\Vtheta$ and $\Vtheta'$ gives the reverse
inclusion. In particular,
\begin{equation}\label{eq:appendix-niv-information}
\sigma\{G(\Vtheta)\}=\sigma(G_0),
\qquad G_0=G(\Vtheta_0).
\end{equation}
We can therefore write the common conditional expectation
\[
\mu_G=\EE[Z\mid G_0]=\EE[Z\mid G(\Vtheta)]
\]
and write $R_Z=Z-\mu_G$.

For fixed $n$, the bounded kernel and fixed scaling factors imply
that $G(\Vtheta)^\ell$ and $J+\delta_0JG_0$ are bounded matrices
for each $\ell=0,\ldots,q$. Hence
\[
\EE[G(\Vtheta)^\ell R_Z\mid G(\Vtheta)]
=
G(\Vtheta)^\ell
\{\EE[Z\mid G(\Vtheta)]-\mu_G\}
=
\boldsymbol0.
\]
Stacking these equalities yields
\begin{equation}\label{eq:appendix-niv-centering}
\EE[\TZ_{\Vtheta}\mid G(\Vtheta)]=\boldsymbol0.
\end{equation}

Moreover, \eqref{eq:appendix-niv-information} implies
$\sigma\{Z,G(\Vtheta)\}=\sigma(Z,G_0)$. Using
$\TVep=(J+\delta_0JG_0)\Vep$ and
Assumption~\ref{assumption: conditional exogeneity}, we obtain
\begin{align}
\EE[\TVep\mid Z,G(\Vtheta)]
&=(J+\delta_0JG_0)\EE[\Vep\mid Z,G_0]\notag\\
&=(J+\delta_0JG_0)\EE[\Vep\mid G_0]\notag\\
&=\EE[\TVep\mid G(\Vtheta)].
\label{eq:appendix-niv-exclusion}
\end{align}
Here $G_0$ is measurable with respect to $\sigma\{G(\Vtheta)\}$,
which justifies the first and last equalities.

The matrix $\TZ_{\Vtheta}$ is measurable with respect to
$\sigma\{Z,G(\Vtheta)\}$. Therefore,
\begin{align}
\EE[\TZ_{\Vtheta}^{\top}\TVep\mid Z,G(\Vtheta)]
&=\TZ_{\Vtheta}^{\top}
  \EE[\TVep\mid Z,G(\Vtheta)]\notag\\
&=\TZ_{\Vtheta}^{\top}
  \EE[\TVep\mid G(\Vtheta)].
\label{eq:appendix-niv-inner-expectation}
\end{align}
If $\TZ_{\Vtheta}$ is unbounded, the first equality follows by
establishing it on
$\{\|\TZ_{\Vtheta}\|_{\mathrm F}\le m\}$ and letting $m$ increase
to infinity. In particular, the final product in
\eqref{eq:appendix-niv-inner-expectation} is integrable because
it equals a conditional expectation of the integrable NIV moment.

Applying the tower property and
\eqref{eq:appendix-niv-centering} now gives
\begin{align*}
\EE[\TZ_{\Vtheta}^{\top}\TVep\mid G(\Vtheta)]
&=
\EE\!\left[
  \EE[\TZ_{\Vtheta}^{\top}\TVep\mid Z,G(\Vtheta)]
  \,\middle|\,G(\Vtheta)
\right]\\
&=
\EE\!\left[
  \TZ_{\Vtheta}^{\top}\EE[\TVep\mid G(\Vtheta)]
  \,\middle|\,G(\Vtheta)
\right]\\
&=
\EE[\TZ_{\Vtheta}\mid G(\Vtheta)]^{\top}
\EE[\TVep\mid G(\Vtheta)]\\
&=\boldsymbol0.
\end{align*}
The third equality is justified, if necessary, by first
restricting to
$\{\|\EE[\TVep\mid G(\Vtheta)]\|\le m\}$ and then letting
$m$ increase to infinity. Finally, taking unconditional
expectations yields
\[
\EE[\TZ_{\Vtheta}^{\top}\TVep]
=
\EE\!\left[
  \EE[\TZ_{\Vtheta}^{\top}\TVep\mid G(\Vtheta)]
\right]
=
\boldsymbol0,
\]
as required.
\end{proof}



\begin{proof}[Proof of Lemma~\ref{lemma: modal extraction}]
Fix $(\TVbeta,\Vtheta)$ and write $t_s=\lambda_s(\Vtheta)$ and
$e=\TVy-\TX(\Vtheta)\TVbeta$. Since $JU_s=U_s$ and both $J$ and
$G(\Vtheta)$ are symmetric, $U_s^\top J=U_s^\top$ and
$U_s^\top G(\Vtheta)=t_sU_s^\top$. Projecting the candidate residual
onto the columns of $U_s$ therefore gives
\[
  U_s^\top e
  =(1-\rho t_s)U_s^\top\Vy
   -U_s^\top X(\Vbeta_1+t_s\Vbeta_2).
\]

Let $m_h$ denote the block of NIV moments associated with the
instrument $G(\Vtheta)^hR_Z$, for $h=0,\ldots,q$. If
$R_Z=\sum_{s=1}^H P_sR_Z$, then
$G(\Vtheta)^hR_Z=\sum_{s=1}^H t_s^hP_sR_Z$.
If instead $\sum_{s=1}^H P_s=J$, then $Je=e$ implies
$G(\Vtheta)^he=\sum_{s=1}^H t_s^hP_se$. Under either condition,
symmetry of $G(\Vtheta)$ and the fact that each $t_s$ is nonrandom
give
\begin{align}
  m_h
  &=\EE\!\left[\frac{1}{n}(G(\Vtheta)^hR_Z)^\top e\right]
    =\sum_{s=1}^H t_s^h
      \EE\!\left[\frac{1}{n}(U_s^\top R_Z)^\top U_s^\top e\right]
      \notag\\
  &=\sum_{s=1}^H t_s^h C_s
    \bigl\{(1-\rho t_s)f_s-\Vbeta_1-t_s\Vbeta_2\bigr\}.
    \label{eq:proof-modal-extraction}
\end{align}
For $h=0,\ldots,H-1$, the matrix of weights $(t_s^h)_{h,s}$ is
Vandermonde with determinant $\prod_{s<u}(t_u-t_s)\neq0$.
Since $q\ge H-1$ and each $C_s$ is nonsingular, vanishing of these
first $H$ moment blocks implies
\[
  (1-\rho t_s)f_s=\Vbeta_1+t_s\Vbeta_2,
  \qquad s=1,\ldots,H.
\]
Conversely, these equations make every block in
\eqref{eq:proof-modal-extraction} zero. This proves
\eqref{eq:modal-equations-new}. Proposition~\ref{Proposition: Valid Moment Condition}
and the structural equation give
$m(\TVbeta_0,\Vtheta_0)=\boldsymbol 0$. Evaluating
\eqref{eq:modal-equations-new} at the truth and dividing by
$1-\rho_0\lambda_s(\Vtheta_0)$ whenever it is nonzero yields
\eqref{eq:modal-response-new}.
\end{proof}

\begin{proof}[Proof of Theorem~\ref{Theorem: Spectral Identification}]
Fix a coordinate $j$ with
$c_{0j}=\beta_{20,j}+\rho_0\beta_{10,j}\neq0$, which exists by
\eqref{eq:noncancellation-new}. For any solution
$(\TVbeta,\Vtheta)\in\RR^p\times\Theta$ of the NIV moments, write
$t_s=\lambda_s(\Vtheta)$. By Lemma~\ref{lemma: modal extraction},
$(1-\rho t_s)f_{s,j}=\beta_{1j}+t_s\beta_{2j}$ for every $s$.
Multiplying the equations for $s$ and $u$ by $1-\rho t_u$ and
$1-\rho t_s$, respectively, and subtracting gives
\begin{equation}\label{eq:eigenvalue-pair-difference}
  (1-\rho t_s)(1-\rho t_u)(f_{s,j}-f_{u,j})
  =(\beta_{2j}+\rho\beta_{1j})(t_s-t_u).
\end{equation}

At the true parameter, the right-hand side of
\eqref{eq:eigenvalue-pair-difference} is nonzero for every $s\neq u$.
Hence the $f_{s,j}$ are pairwise distinct. Now consider any candidate
solution. Since the candidate eigenvalues are distinct, at most one
factor $1-\rho t_s$ can vanish. Choose two indices for which both
factors are nonzero. Their values of $f_{s,j}$ differ, so
\eqref{eq:eigenvalue-pair-difference} implies
$\beta_{2j}+\rho\beta_{1j}\neq0$. Applying the same identity to every
pair then shows that $1-\rho t_s\neq0$ for every $s$.

For each group $r$, substitute \eqref{eq:eigenvalue-pair-difference}
into the four differences defining the cross-ratio in
\eqref{eq:cross-ratio-new}. The common coefficient
$\beta_{2j}+\rho\beta_{1j}$ cancels, and each of the four factors
$1-\rho\lambda_{r\ell}(\theta_r)$ occurs once in the numerator and
once in the denominator. Thus every solution satisfies
\[
  \chi_r(\theta_r)
  =\mathcal C(f_{r1,j},f_{r2,j},f_{r3,j},f_{r4,j})
  =\chi_r(\theta_{0r}),
\]
where the last equality follows by applying the same identity at the
truth. Injectivity of $\chi_r$ gives \eqref{eq:theta-recovery-new}
and hence $\Vtheta=\Vtheta_0$.

It remains to identify the structural coefficients. Take three
distinct eigenvalues $t_1,t_2,t_3$ at $\Vtheta_0$. Applying
\eqref{eq:eigenvalue-pair-difference} to the pairs $(1,2)$ and
$(1,3)$, first at the candidate and then at the truth, yields
\[
  \frac{1-\rho t_3}{1-\rho t_2}
  =
  \frac{(f_{1,j}-f_{2,j})/(t_1-t_2)}
       {(f_{1,j}-f_{3,j})/(t_1-t_3)}
  =
  \frac{1-\rho_0 t_3}{1-\rho_0 t_2}.
\]
Cross-multiplication gives $(\rho-\rho_0)(t_2-t_3)=0$, so
$\rho=\rho_0$. Subtracting the true and candidate vector equations
in \eqref{eq:modal-equations-new} now gives
\[
  (\Vbeta_1-\Vbeta_{10})+t_s(\Vbeta_2-\Vbeta_{20})
  =\boldsymbol 0,\qquad s=1,2.
\]
Since $t_1\neq t_2$, subtraction yields $\Vbeta_2=\Vbeta_{20}$,
and substitution gives $\Vbeta_1=\Vbeta_{10}$.

Thus $(\TVbeta_0,\Vtheta_0)$ is the unique zero on
$\RR^p\times\Theta$. At $\Vtheta_0$, uniqueness of the solution to
the linear system $M(\Vtheta_0)\TVbeta=\Vc_{\Vtheta_0}$ implies
$\operatorname{rank}M(\Vtheta_0)=p$. For any
$\Vtheta\neq\Vtheta_0$, the moment decomposition
\eqref{eq:moment-decomposition-new} implies that
$d(\Vtheta)\in\operatorname{col}M(\Vtheta)$ would yield a solution
at that $\Vtheta$. Hence $\Pi_{\Vtheta}^{\perp}d(\Vtheta)\neq
\boldsymbol 0$ for every $\Vtheta\neq\Vtheta_0$, as required by
Assumption~\ref{assumption: high level separation}.
\end{proof}



Fix a group $r$ and its observed distances. Let
$c_{r1},\ldots,c_{rL_r}$ be the distinct distances among pairs of
different units, including zero if two different units have distance
zero. For each $h$, define the $n_r\times n_r$ matrix $D_{rh}$ by
\[
  (D_{rh})_{ij}=\mathbb I\{i\neq j,\ d_{ij}=c_{rh}\}.
\]
Under the exponential kernel, the candidate network matrix is
\begin{equation}\label{eq:exponential-distance-decomposition}
  G_r(t)=a_{r,n}\sum_{h=1}^{L_r}e^{-t c_{rh}}D_{rh},
  \qquad t\in\Theta_r.
\end{equation}

\begin{lemma}[Common eigenvectors under the exponential kernel]
\label{lemma: distance classes}
Suppose $\Theta_r$ contains an open interval. A fixed unit vector
$u\in\RR^{n_r}$ is an eigenvector of $G_r(t)$ for every
$t\in\Theta_r$ if and only if it is an eigenvector of every $D_{rh}$.
In that case, its eigenvalue under $G_r(t)$ is
\begin{equation}\label{eq:eigenvalue-from-distances}
  \lambda_r(t)
  =a_{r,n}\sum_{h=1}^{L_r}\gamma_{rh}e^{-t c_{rh}},
  \qquad
  \gamma_{rh}=u^\top D_{rh}u.
\end{equation}
\end{lemma}

\begin{proof}[Proof of Lemma~\ref{lemma: distance classes}]
If $D_{rh}u=\gamma_{rh}u$ for every $h$, multiplying
\eqref{eq:exponential-distance-decomposition} by $u$ gives
$G_r(t)u=\lambda_r(t)u$ with $\lambda_r(t)$ as in
\eqref{eq:eigenvalue-from-distances}.

Now suppose $G_r(t)u\in\operatorname{span}\{u\}$ for every
$t\in\Theta_r$. Choose $L_r$ equally spaced points
$t_j=t_*+j\Delta$, $j=0,\ldots,L_r-1$, in the stated interval,
with $\Delta>0$. Evaluating
\eqref{eq:exponential-distance-decomposition} at these points gives
\[
  a_{r,n}^{-1}G_r(t_j)u
  =\sum_{h=1}^{L_r}
    (e^{-\Delta c_{rh}})^j e^{-t_*c_{rh}}D_{rh}u,
  \qquad j=0,\ldots,L_r-1.
\]
The coefficient matrix on the right is a Vandermonde matrix in the
distinct numbers $e^{-\Delta c_{rh}}$ multiplied by a diagonal
matrix with positive entries $e^{-t_*c_{rh}}$. It is therefore
invertible. Then each $D_{rh}u$ is a linear combination of
the vectors $G_r(t_j)u$, all of which belong to
$\operatorname{span}\{u\}$. Thus $D_{rh}u=\gamma_{rh}u$ for every
$h$. Premultiplication by $u^\top$ gives
$\gamma_{rh}=u^\top D_{rh}u$ because $u^\top u=1$.
\end{proof}

To obtain the group-specific eigenvectors used in
Assumption~\ref{assumption: extractable modes}, select an eigenvalue
$\gamma_{rh}$ of each $D_{rh}$ and form the intersection
\[
  \boldsymbol 1_{n_r}^{\perp}
  \;\cap\;
  \bigcap_{h=1}^{L_r}
  \ker\!\left(D_{rh}-\gamma_{rh}I_{n_r}\right).
\]
Every vector $u$ in this intersection satisfies $J_ru=u$. If the
intersection is nonzero, orthonormal vectors from it, extended by
zeros outside group $r$, give columns of a matrix $U_s$ in
\eqref{eq:common-modes-new}. All these columns share the eigenvalue
function in \eqref{eq:eigenvalue-from-distances}. This formula gives the functions needed
to check that the selected eigenvalues are distinct at every candidate
$\Vtheta$ and that the four labeled eigenvalues in each group yield a
one-to-one cross-ratio $\chi_r$ on $\Theta_r$.



\begin{proof}[Proof of Theorem~\ref{Theorem: identification under span restriction}]
Proposition~\ref{Proposition: Valid Moment Condition} implies
$m(\TVbeta_0,\Vtheta_0)=\boldsymbol0$. Consider any
$(\TVbeta,\Vtheta)\in\RR^p\times\Theta$ satisfying
$m(\TVbeta,\Vtheta)=\boldsymbol0$.

Applying the candidate NIV to the true structural equation and
using Proposition~\ref{Proposition: Valid Moment Condition} gives
\[
\Vc_{\Vtheta}
=
A_{\Vtheta}\Vbeta_{10}
+
B_{\Vtheta}(\Vtheta_0)\Vgamma_0.
\]
The candidate moment equation also implies
\[
\Vc_{\Vtheta}
=
M_{\Vtheta}(\Vtheta)\TVbeta
=
A_{\Vtheta}\Vbeta_1
+
B_{\Vtheta}(\Vtheta)
\begin{pmatrix}
\rho\\
\Vbeta_2
\end{pmatrix}.
\]
Subtracting these representations yields
\[
B_{\Vtheta}(\Vtheta_0)\Vgamma_0
=
A_{\Vtheta}(\Vbeta_1-\Vbeta_{10})
+
B_{\Vtheta}(\Vtheta)
\begin{pmatrix}
\rho\\
\Vbeta_2
\end{pmatrix}
\in
\operatorname{Span}
\{A_{\Vtheta},B_{\Vtheta}(\Vtheta)\}.
\]
Assumption~\ref{assumption: global separation universal}
rules out this inclusion whenever $\Vtheta\ne\Vtheta_0$.
Hence $\Vtheta=\Vtheta_0$.

At the true network parameter, the true and candidate moment
equations imply
\[
M_{\Vtheta_0}(\Vtheta_0)(\TVbeta-\TVbeta_0)
=
\boldsymbol0.
\]
By Assumption~\ref{assumption: rank conditions}(ii),
$M_{\Vtheta_0}(\Vtheta_0)$ has full column rank $p = 2k+1$,
so $\TVbeta=\TVbeta_0$. Therefore,
$(\TVbeta_0,\Vtheta_0)$ is the unique solution of the NIV
moment equations.
\end{proof}



\begin{proof}[Proof of Theorem~\ref{Theorem: disturbance identification}]
Once $(\Vtheta_0,\TVbeta_0)$ are identified, the structural residual
\[
\boldsymbol{u}_0
=
\TVy-\TX\TVbeta_0
=
(J+\delta_0JG_0)\Vep
\]
is identified, with $G_0\equiv G(\Vtheta_0)$. Hence its conditional covariance matrix is identified:
$$
\cov(\boldsymbol{u}_0\mid G_0)=\sigma_{\ep}^{2}\Sigma(\delta_0,\Vtheta_0).
$$

Suppose another pair $(\delta,\sigma^{2})$ generates the same covariance matrix:
\[
\sigma^{2}\Sigma(\delta,\Vtheta_0)=\sigma_{\ep}^{2}\Sigma(\delta_0,\Vtheta_0).
\]
If $\delta\neq \delta_0$, then
\[
\Sigma(\delta,\Vtheta_0)=\frac{\sigma_{\ep}^{2}}{\sigma^{2}}\Sigma(\delta_0,\Vtheta_0),
\]
which contradicts Assumption \ref{assumption: covariance shape}. Therefore $\delta=\delta_0$. Substituting back gives
\[
\sigma^{2}\Sigma(\delta_0,\Vtheta_0)=\sigma_{\ep}^{2}\Sigma(\delta_0,\Vtheta_0).
\]
Since $\Sigma(\delta_0,\Vtheta_0)\neq \boldsymbol{0}$ whenever at least one group has size greater than one, it follows that $\sigma^{2}=\sigma_{\ep}^{2}$. Hence $(\delta_0,\sigma_{\ep}^{2})$ are uniquely identified.
\end{proof}

\begin{proof}[Proof of Theorem~\ref{thm:joint-gmm}]
Let $Q(\eta)=m(\eta)^\top Wm(\eta)$. Assumption~\ref{ass:joint-consistency}
gives uniform convergence of $\hat m_N$ to $m$ and $W_N\to_p W$.
Since $m$ is bounded on $\mathcal H$,
\begin{align*}
 \sup_\eta|\hat Q_N(\eta)-Q(\eta)|
 &\le \|W_N\|\sup_\eta\|\hat m_N(\eta)-m(\eta)\|
       \left\{\sup_\eta\|\hat m_N(\eta)\|+\sup_\eta\|m(\eta)\|\right\}\\
 &\quad+\|W_N-W\|\sup_\eta\|m(\eta)\|^2=o_p(1).
\end{align*}
Since $Q(\eta_0)=0$, approximate minimization implies
$0\le Q(\hat\eta)\le2\sup_\eta|\hat Q_N(\eta)-Q(\eta)|+o_p(1)$.
For each $\epsilon>0$ with a nonempty complement of the
$\epsilon$-ball in $\mathcal H$, continuity, compactness, and uniqueness give
$\inf_{\|\eta-\eta_0\|\ge\epsilon}Q(\eta)>0$.
It follows that $\hat\eta\to_p\eta_0$, as in the extremum-estimation
argument of \citet{NeweyMcFadden1994JointNote}.

For asymptotic normality, we first bound the error from estimating the
conditional means. With row and power indices omitted,
$\EE[Z-\mu\mid\mathcal F_G]=\EE[w-a\mid\mathcal F_G]=0$. Thus, for integrable
$\mathcal F_G$-measurable perturbations $h_\mu,h_a$,
\[
 \left.\frac{d}{dt}\EE[(Z-\mu-th_\mu)^\top(w-a-th_a)]\right|_{t=0}
 =-\EE[h_\mu^\top(w-a)]-\EE[(Z-\mu)^\top h_a]=0.
\]
Expanding the product at the estimated conditional means gives
\eqref{eq:joint-remainder}. The norm of the last term is at most
$r_{\mu,N}r_{a,N}$ by Cauchy--Schwarz. Assumption~\ref{ass:joint-regular}(ii)
therefore implies
\[
 \sqrt N\{\hat m_N(\eta_0)-m_N^0(\eta_0)\}=o_p(1).
\]
Applying the same bounds uniformly gives
$\sup_\eta\|\hat m_N(\eta)-m_N^0(\eta)\|=o_p(1)$ when the two empirical
$L^2$ nuisance errors are uniformly $o_p(1)$ and the empirical second
moments of $Z-\mu$ and $w(\eta)-a(\eta)$ are uniformly $O_p(1)$.
An oracle uniform law then gives the uniform moment convergence in
Assumption~\ref{ass:joint-consistency}(ii).

Write $\hat D=\partial_\eta\hat m_N(\hat\eta)$ and
\[
 \bar D_N=\int_0^1\partial_\eta\hat m_N
                 \{\eta_0+t(\hat\eta-\eta_0)\}\,dt.
\]
Consistency and Assumption~\ref{ass:joint-regular}(i) give
$\hat D\to_p D$ and $\bar D_N\to_p D$. Expanding only
$\hat m_N(\hat\eta)$ in the first-order condition gives
\[
 \hat D^\top W_N\bar D_N(\hat\eta-\eta_0)
 =-\hat D^\top W_N\hat m_N(\eta_0)+o_p(N^{-1/2}).
\]
Since the coefficient matrix converges to $D^\top WD$, which is
nonsingular,
\begin{align*}
 \sqrt N(\hat\eta-\eta_0)
 &=-\{\hat D^\top W_N\bar D_N\}^{-1}\hat D^\top W_N
                    \sqrt N\hat m_N(\eta_0)+o_p(1)\\
 &=-A\sqrt N m_N^0(\eta_0)+o_p(1)
   \Rightarrow\mathcal N(0,A\Omega A^\top).
\end{align*}

Let $\dot G_r=\partial_{\theta_r}G(\Vtheta)$, embedded in the full
block-diagonal matrix. The product rule gives
\begin{equation}\label{eq:joint-power-derivative}
 \partial_{\theta_r}G(\Vtheta)^s
   =\sum_{h=0}^{s-1}G(\Vtheta)^h\dot G_rG(\Vtheta)^{s-1-h},
 \qquad s\ge1.
\end{equation}
Put $T_0=\rho_0\Vy+X\Vbeta_{20}$ and
$M_N(\Vtheta)=\EE[N^{-1}\TZ_{\Vtheta}^\top J R_x(\Vtheta)]$.
When differentiation under the expectation is valid,
$D_N=\partial_\eta m_N(\eta_0)$ has blocks
\begin{align}
 D_{\beta,N}&=-M_N(\Vtheta_0),\label{eq:joint-beta-derivative}\\
 [D_{\theta,N}]_{s,r}
   &=-\EE\left[\frac1N R_Z^\top(G_0^s)^\top J\dot G_{0r}T_0\right],
      \qquad s=0,\ldots,q.
      \label{eq:joint-theta-derivative}
\end{align}
Here $[D_{\theta,N}]_{s,r}$ is row block $s$ of column $r$.
To see this, differentiate the $s$th block with respect to $\theta_r$:
\[
 \EE\left[\frac1N R_Z^\top
       \{\partial_{\theta_r}G(\Vtheta)^s\}^\top u(\eta)\right]
 -\EE\left[\frac1N R_Z^\top(G(\Vtheta)^s)^\top
                        J\dot G_r(\rho\Vy+X\Vbeta_2)\right].
\]
At $\eta_0$, conditional exclusion gives, for each component $a$,
\[
 \frac1N\sum_{i,j}
 \left[\{\partial_{\theta_r}G(\Vtheta_0)^s\}^\top
                      J(I_N+\delta_0G_0)\right]_{ij}
 \EE[(Z_{ia}-\mu_{ia})\varepsilon_j\mid\mathcal F_G]=0.
\]
Thus the first term vanishes, yielding \eqref{eq:joint-theta-derivative}.
Linearity in $\TVbeta$ gives \eqref{eq:joint-beta-derivative}. The
augmented moments have the same Jacobian because their population
values equal those of the original moments.

If the weight depends on $\eta$, use
$Q(\eta)=m(\eta)^\top W(\eta)m(\eta)$ in the consistency argument.
For a consistent approximate stationary point with gradient residual
$o_p(N^{-1/2})$, let $x_N=\|\hat\eta-\eta_0\|$. The local derivative
bounds imply
\begin{align*}
 \|\hat m_N(\hat\eta)\|&=O_p(N^{-1/2}+x_N),\\
 \max_j\left|\hat m_N(\hat\eta)^\top
      \partial_{\eta_j}W_N(\hat\eta)\hat m_N(\hat\eta)\right|
      &=O_p(N^{-1}+x_N^2).
\end{align*}
The first-order condition therefore gives, for some nonnegative
$C_N=O_p(1)$,
\[
 x_N\le C_N(N^{-1/2}+x_N^2),\qquad
 x_N(1-C_Nx_N)\le C_NN^{-1/2}.
\]
Since $C_Nx_N=o_p(1)$, $x_N=O_p(N^{-1/2})$.
The additional gradient is then $O_p(N^{-1})$, giving the same limit
with $W=W(\eta_0)$.

For covariance estimation, put
\[
 \tilde\Omega_N=\frac1N\sum_{i,j}\omega_{N,ij}\psi_i\psi_j^\top.
\]
For each entry $(a,b)$ and $\epsilon>0$, conditional Chebyshev gives
\begin{align*}
 &\Pr\left(\left|[\tilde\Omega_N
                   -\EE(\tilde\Omega_N\mid\mathcal F_G)]_{ab}\right|
                       >\epsilon\right)\\
 &\qquad\le\EE\left[\min\left\{1,\epsilon^{-2}
          \operatorname{Var}([\tilde\Omega_N]_{ab}\mid\mathcal F_G)
                       \right\}\right]\to0
\end{align*}
by \eqref{eq:joint-hac-variance} and boundedness of the integrand.
Equations~\eqref{eq:joint-hac-bias} and \eqref{eq:joint-hac-limit} give
$\EE(\tilde\Omega_N\mid\mathcal F_G)\to_p\Omega$, so
$\tilde\Omega_N\to_p\Omega$. Symmetry of $\omega_{N,ij}$ and
Cauchy--Schwarz yield
\[
 \|\hat\Omega_N-\tilde\Omega_N\|
 \le B_N\left\{2\left(\frac1N\sum_i\|\psi_i\|^2\right)^{1/2}e_N
                   +e_N^2\right\}=o_p(1)
\]
by \eqref{eq:joint-hac-limit}--\eqref{eq:joint-hac-plugin}.
It follows that $\hat\Omega_N\to_p\Omega$, $\hat A_N\to_p A$, and
$\hat V_N\to_p V$. For a regular restriction $h$, the delta method gives
$\sqrt N h(\hat\eta)\Rightarrow\mathcal N(0,H_0VH_0^\top)$.
Since $\hat H\hat V_N\hat H^\top\to_p H_0VH_0^\top\succ0$,
studentization gives \eqref{eq:joint-wald} and the marginal intervals.
Inverting $D^\top\Omega^{-1}D$ by blocks gives the coefficient covariance.

As one possible fixed-shape procedure, for each $a\in\mathcal A$,
compute the identity-weight estimate
\[
 \check\eta(a)\in\arg\min_{\eta\in\mathcal H}
                      \|\hat m_N(\eta;a)\|^2.
\]
With the true shape held fixed, Assumptions~\ref{ass:joint-consistency}--%
\ref{ass:joint-regular} for the remaining parameters give
$\check\eta(a_0)-\eta_0=O_p(N^{-1/2})$ by Theorem~\ref{thm:joint-gmm}.
Estimate the covariance at the initial fit to obtain $W_N$ for the
one-step update \citep[Section~3.4]{NeweyMcFadden1994JointNote}.
Suppressing the fixed shape argument, suppose
$\|\check\eta-\eta_0\|=o_p(N^{-1/4})$ and the feasible Jacobian is locally
Lipschitz with an $O_p(1)$ Lipschitz bound. The one-step estimate is
\begin{equation}\label{eq:joint-one-step}
 \eta^{\rm os}
 =\check\eta-
 \{\hat D(\check\eta)^\top W_N\hat D(\check\eta)\}^{-1}
       \hat D(\check\eta)^\top W_N\hat m_N(\check\eta).
\end{equation}
At the true fixed shape, put $h_N=\check\eta-\eta_0$ and
$\hat A(\check\eta)=
\{\hat D(\check\eta)^\top W_N\hat D(\check\eta)\}^{-1}
\hat D(\check\eta)^\top W_N$.
By the local Lipschitz condition,
$\hat m_N(\check\eta)=\hat m_N(\eta_0)+\hat D(\check\eta)h_N
+O_p(\|h_N\|^2)$. At $\check\eta$ the weight and derivative converge
to $W$ and $D$, so $\hat A(\check\eta)\to_p A$. Also
$\sqrt N\|h_N\|^2=(N^{1/4}\|h_N\|)^2=o_p(1)$.
In \eqref{eq:joint-one-step}, $\hat A(\check\eta)\hat D(\check\eta)=I$,
which cancels the term in $h_N$. This leaves
\[
 \sqrt N(\eta^{\rm os}-\eta_0)
 =-\hat A(\check\eta)\sqrt N\hat m_N(\eta_0)
       +O_p(\sqrt N\|h_N\|^2)
 =-A\sqrt N m_N^0(\eta_0)+o_p(1).
\]
Under the full-score alternative \eqref{eq:joint-full-score}, the same cancellation instead gives $\sqrt N(\eta^{\rm os}-\eta_0)=-A N^{-1/2}\sum_i\psi_i^{\mathrm{full}}+o_p(1)$; the stated CLT and feasible covariance conditions must hold for that full score and its estimator.

Write $I_j(a)=[L_j(a),U_j(a)]$ for the interval for coefficient
$\beta_{0j}$ of $(\Vbeta_{10}^\top,\Vbeta_{20}^\top)^\top$ when we hold
the shape at $a$. The smallest interval containing the grid intervals is
\[
 I_j^{\rm env}
   =\left[\min_{a\in\mathcal A}L_j(a),\
           \max_{a\in\mathcal A}U_j(a)\right].
\]
Since $a_0\in\mathcal A$, the interval at $a_0$ is one of the intervals
used in both extrema. The probability of coverage can fall below that
of $I_j(a_0)$ by at most the probability of a missing grid interval,
which is $o(1)$. Hence
\[
 \liminf_{N\to\infty}\Pr\{\beta_{0j}\in I_j^{\rm env}\}
 \ge \lim_{N\to\infty}\Pr\{\beta_{0j}\in I_j(a_0)\}
 =1-\alpha.
\]

For the score CLT, let $B_{Nm}$ be independent centered block vectors,
whose number increases while $\bar r$ remains fixed, and suppose
\[
 \begin{gathered}
 N^{-1/2}\left\|\sum_i\psi_i-\sum_mB_{Nm}\right\|=o_p(1),\qquad
 N^{-1}\sum_m\operatorname{Var}(B_{Nm})\to\Omega,\\
 N^{-1-\xi/2}\sum_m\EE\|B_{Nm}\|^{2+\xi}\to0.
 \end{gathered}
\]
for some $\xi>0$. To apply Lyapunov's theorem, project the blocks
onto $v$ and let $s_N^2=\sum_m\operatorname{Var}(v^\top B_{Nm})$.
If $v^\top\Omega v>0$, the limit of $s_N^2/N$ is positive and
\[
 \frac{\sum_m\EE|v^\top B_{Nm}|^{2+\xi}}{s_N^{2+\xi}}
 \le\frac{\|v\|^{2+\xi}N^{-1-\xi/2}
                  \sum_m\EE\|B_{Nm}\|^{2+\xi}}
                 {(s_N^2/N)^{1+\xi/2}}\to0.
\]
For the zero-variance projections, Chebyshev's inequality with
variance $s_N^2/N$ gives the degenerate limit. The Cram\'er--Wold
device now gives the multivariate block limit, and the first bound
transfers it to the score. These bounds can be imposed on the
adjusted scores, or under the conditional law for a conditional CLT.
\end{proof}

\begin{proof}[Proof of Theorem~\ref{thm:restricted-niv}]
Put $b_0=\Vbeta_{10}$. Under $H_0$,
$J(\Vy-Xb_0)=J(I_N+\delta_0G_0)\Vep$.
For every $s$, $\{G(\bar{\Vtheta})^s\}^\top J(I_N+\delta_0G_0)$ is
$\mathcal F_G$-measurable, so conditional exclusion implies
\[
 \EE[R_Z^\top\{G(\bar{\Vtheta})^s\}^\top J(\Vy-Xb_0)]=0.
\]
The equality holds for any reference $\bar{\Vtheta}$.
Subtracting a conditional mean from the transformed response adds a
term with expectation zero since $\EE[R_Z\mid\mathcal F_G]=0$.
By Assumption~\ref{ass:restricted-niv}, $\hat M_0$ is nonsingular with
probability tending to one and
$\hat b-b_0=\hat M_0^{-1}\hat g_0(b_0)=O_p(N^{-1/2})$.
Linearity gives
\[
 \hat g_+(\hat b)
 =\hat g_+(b_0)-\hat M_+\hat M_0^{-1}\hat g_0(b_0),\qquad
 \sqrt N\hat z=C\hat L\sqrt N\hat g(b_0)
       \Rightarrow\mathcal N(0,V_C).
\]
Together with $\hat V_C\to_p V_C\succ0$, this implies
$J_C\Rightarrow\chi^2_d$. Since $\chi^2_{d,1-\alpha}$ is a continuity
point, the rejection probability tends to $\alpha$.
When the assumptions hold conditionally for almost every common design
environment, the conditional rejection probability converges to $\alpha$.
The unconditional probability then converges to $\alpha$ by dominated
convergence.

For any deterministic $b_N$, $\hat z=C\hat L\hat g(b_N)$.
Under the local alternatives,
\[
 \sqrt N\hat z=C\hat L\sqrt N\hat g(b_N)
       \Rightarrow\mathcal N(CLh,V_C),\qquad
 J_C\Rightarrow\chi^2_d\bigl((CLh)^\top V_C^{-1}(CLh)\bigr).
\]
Under a fixed alternative with $z_A\ne0$,
\[
 \frac{J_C}{N}\to_p z_A^\top V_A^{-1}z_A>0,\qquad
 \Pr(J_C>\chi^2_{d,1-\alpha})\to1.
\]

For fold $E$ in group $r$, let $e_i=\mathbf1\{i\in E\}$ and $m_E=|E|$.
Take $F$ to be the matrix of the fixed set of geometry features
used for balancing, including a constant. Assume that the eigenvalues
of $F^\top F/n_r$ are bounded above and away from zero. Define
\begin{equation}\label{eq:restricted-balance}
 w_E=F(F^\top F)^{-1}F^\top e,\qquad F^\top w_E=F^\top e.
\end{equation}
Estimate $a_E$ and $\gamma_{sE}$ on $T_E$. The fitted instrument
and response means at $i\in E$ are $\mu_{ij}(\hat a_E)$ and
$F_i\hat\gamma_{sE}$.
For instrument $j$ and power $s$, put
$V_{si}=v_{rsi}(\Vy_r-X_r\hat b,\bar\theta_r)$ and, with gradients written
as columns, define
\[
 d^Z_{i,js}=\dot\mu_{ij}(\hat a_E)(V_{si}-F_i\hat\gamma_{sE}),\qquad
 d^V_{i,j}=F_i^\top\{Z_{rij}-\mu_{ij}(\hat a_E)\}.
\]
Replace $\sum_i e_i d_i^Z$ and $\sum_i e_i d_i^V$ by
$D_E^Z=\sum_i w_{E,i}d_i^Z$ and $D_E^V=\sum_i w_{E,i}d_i^V$, respectively.

For a regular training estimate $t$ with
$\sum_{u\in T_E}s_u(\hat t)=0$, the estimated influence contribution of
observation $i$ is
\[
 I^t_{iE}
 =-\left\{\sum_{u\in T_E}\partial_t s_u(\hat t)\right\}^{-1}s_i(\hat t)
       \mathbf1\{i\in T_E\}.
\]
Assume a differentiable training equation with nonsingular limiting
Jacobian and a first-order influence expansion for the estimate
\citep[Section~6]{NeweyMcFadden1994JointNote}. Suppressing the instrument
and power subscripts on the fits, the adjusted score is
\[
 \hat\psi_{i,js}^{\rm bal}
 =\hat\psi_{i,js}^{\rm raw}
       -\sum_E\{(D_E^Z)^\top I^a_{iE}
                         +(D_E^V)^\top I^\gamma_{iE}\}.
\]
Here $\hat\psi^{\rm raw}$ is the product of the held-out instrument and
response residuals. Use the adjusted score in \eqref{eq:joint-hac}.

Take the unbalanced covariance to be consistent under
Assumption~\ref{ass:restricted-niv}. As $N$ increases, the counts of
groups, folds, geometry features and derivative entries do not change.
The empirical second moments of the unbalanced projected scores and of
$m_E I^t_{iE}$ must be $O_p(1)$. To measure the change in the
derivative sums, use
\begin{equation}\label{eq:restricted-balance-rate}
 r_{B,N}=\max_{r,E,\text{ entries},\,t=Z,V}
       \frac1{m_E}\left|\sum_i(w_{E,i}-e_i)d_i^t\right|.
\end{equation}
Suppose $B_N\ge1$ and $B_Nr_{B,N}=o_p(1)$.

Write $A$ and $A+\Delta$ for the projected influence matrices before
and after balancing. Each derivative-sum difference divided by $m_E$
is bounded by $r_{B,N}$ in absolute value. In each row of $\Delta$
these ratios multiply the terms $m_E I^t_{iE}$. Thus the second-moment
bounds and the fixed number of terms per row give
\[
 \frac{\|A\|_F}{\sqrt N}=O_p(1),\qquad
 \frac{\|\Delta\|_F}{\sqrt N}=O_p(r_{B,N}).
\]
Hence
\[
 \|\hat V_{C,\rm bal}-\hat V_C\|
 \le\frac{B_N}{N}(2\|A\|_F\|\Delta\|_F+\|\Delta\|_F^2)
 =O_p(B_Nr_{B,N}+B_Nr_{B,N}^2)=o_p(1),
\]
because $B_N\ge1$ and $B_Nr_{B,N}=o_p(1)$.
Both covariance estimators tend to $V_C\succ0$ and use the same
test numerator. They therefore give the same null limit.

Consider a group of size $n$. Cycle through a fixed number of folds,
assigning a block of $l_n=o(n)$ consecutive units to each fold in turn.
Suppose $m_E/n$ and the smallest eigenvalue of $F^\top F/n$ are bounded
away from zero. In the observed order, the columns of $F$ and
$\bar d_i(t)=\EE[d_i(t)\mid\mathcal F_G]$ have bounded sup norm and total
variation uniformly over $t\in\mathcal T$. Here
$\mathcal T\subset\RR^v$ is compact, $v$ is fixed, and all fitted
arguments lie in $\mathcal T$ with probability tending to one.
Assume for $t\in\mathcal T$ that
\[
 A_n(t)=\frac1{m_E}\sum_i(w_{E,i}-e_i)\{d_i(t)-\bar d_i(t)\}
       =c_n(t)^\top\xi_n,\qquad
 \sup_{t\in\mathcal T}\|c_n(t)\|_2\le Cn^{-1/2},
\]
where $c_n(t)$ is $\mathcal F_G$-measurable and, conditionally on
$\mathcal F_G$, the components of $\xi_n$ are independent, centered,
and uniformly sub-Gaussian. Take the Lipschitz constant of $A_n$ on
$\mathcal T$ to be $O_p(1)$.

Set $p_n=m_E/n$ and $S_t=\sum_{i\le t}(e_i-p_n)$.
If there are $L$ folds, $\sum_{i\le t}e_i=t/L+O(l_n)$ uniformly in $t$,
because only the last cycle can be incomplete. Taking $t=n$ gives
$p_n=1/L+O(l_n/n)$. Thus $\max_t|S_t|\le Cl_n$ and $S_n=0$.
Summation by parts gives
\[
 \left|\sum_i(e_i-p_n)f_i\right|
 \le Cl_n\{\|f\|_\infty+\operatorname{TV}(f)\}.
\]
Since $F$ contains an intercept,
$w_E-p_n\mathbf1=F(F^\top F/n)^{-1}
\{F^\top(e-p_n\mathbf1)/n\}$, and hence
$\max_i|w_{E,i}-p_n|=O(l_n/n)$.
Applying the preceding bounds to $\bar d_i(t)$ yields, uniformly in $t$,
\[
 \left|\frac1{m_E}\sum_i(w_{E,i}-e_i)\bar d_i(t)\right|=O(l_n/n).
\]
For each $t$, $\Pr\{|A_n(t)|>x\mid\mathcal F_G\}\le2\exp(-cnx^2)$ by
sub-Gaussianity. For an $n^{-2}$-net $\mathcal T_n$ of $\mathcal T$,
the conditional probability that $\max_{t\in\mathcal T_n}|A_n(t)|$
exceeds $M\sqrt{\log n/n}$ is bounded by
$2|\mathcal T_n|n^{-cM^2}=O(n^{2v-cM^2})$. Since
$\sup_{t\in\mathcal T}|A_n(t)|\le
\max_{t\in\mathcal T_n}|A_n(t)|+O_p(n^{-2})$ by the Lipschitz condition,
the bound $O_p(\sqrt{\log n/n})$ applies at the fitted arguments.
Since the numbers of groups, folds, and derivative entries are fixed
and group sizes are proportional to $N$,
\[
 r_{B,N}=O_p\{l_n/n+\sqrt{\log n/n}\}.
\]
For $l_n=\max(3,\lfloor n^{0.45}\rfloor)$ and ordered Bartlett weights
with $h_n=\max(2,\lfloor n^{1/4}\rfloor)$, $B_N\le1+h_n$ and
\[
 B_Nr_{B,N}
 =O_p(n^{-0.30}+n^{-0.25}\sqrt{\log n})=o_p(1).
\]
The coefficient bound on $c_n(t)$ follows, for example, when the
centered derivatives have a linear-filter representation with uniformly
bounded operator norm and $\sum_i|w_{E,i}-e_i|^2/n$ is bounded.
\end{proof}

\clearpage
\section{Simulation design and implementation}\label{app:peer_cell}

\paragraph{Data generation.}
We divide group $r$ into $H_r=\lfloor n_r/6\rfloor$ cells. The first
$n_r\bmod6$ cells have seven units and the rest have six. For each cell
we draw $A_{rh}\sim\operatorname{Unif}[0.15,3.5]$ independently. For
$i\ne k$ in cells $h$ and $\ell$ of group $r$, the squared distance is
\[
 d_{ik}=
 \begin{cases}
 A_{rh}, & \text{within a pair},\\
 A_{rh}+6/c_r, & \text{same cell, other units},\\
 12|h-\ell|/c_r+(A_{rh}+A_{r\ell})/2, & h\ne\ell.
 \end{cases}
\]
$G_{0r}$ has a zero diagonal and off-diagonal entries
$K_{c_r}^{\mathrm G}(d_{ik})$ or $K_{c_r}^{\mathrm L}(d_{ik})$.

The five unit-length contrasts $v_{rhj}$ are zero outside the
first six units of the cell. The first three are within-pair differences,
and the last two are Helmert contrasts of the three pair means.
Write $T_{rh}=(A_{rh}-0.15)/3.35$ and let
\[
\begin{aligned}
 s_{rh}&=\sin(2\pi T_{rh}), & f_r&=(r-1)/9,\\
 b_{rj}&=0.60+0.55\{1+\sin(1.31r+1.73j)\},&
 \pi_{rj}&=1.50+0.90\{1+\cos(1.17r+1.91j)\},\\
 p_{rhj}&=\{1+\exp(-b_{rj}s_{rh})\}^{-1},&
 \lambda_j&=0.8-0.1(j-1).
\end{aligned}
\]
Independently for each index, draw $U_{rhj}\sim\operatorname{Unif}[0,1]$,
$w_{rhj}\sim N(0,1)$, and $\zeta_{ri}\sim N(0,1)$. These three sets of draws
are independent of one another and of the cell draws. With
$z_{rhj}=\sqrt{12p_{rhj}(1-p_{rhj})}\allowbreak(U_{rhj}-1/2)$, the instruments,
covariates, and errors are
\[
\begin{aligned}
 Z_{rj}&=\sum_h v_{rhj}(p_{rhj}+z_{rhj}),\\
 X_{rj}&=\sum_h v_{rhj}\{\pi_{rj}z_{rhj}
                 +\lambda_j v_{rhj}^{\top}\zeta_r+0.45w_{rhj}\},\\
 \varepsilon_r&=(0.30+0.08f_r)\sum_{h,j}s_{rh}v_{rhj}+0.35\zeta_r.
\end{aligned}
\]
We set $\alpha_r=-0.30+0.60f_r$ and solve for $Y_r$ from
\[
 (I-\rho_0G_{0r})Y_r=
 X_r\boldsymbol{\beta}_{10}+G_{0r}X_r\boldsymbol{\beta}_{20}
 +\alpha_r\boldsymbol 1+(I+\delta_0G_{0r})\varepsilon_r .
\]
Within each replication we use the same draws under the Gaussian and
logistic kernels. We set the seed to $180800000+10000q+m$ for
replication $m$ at the $q$th group size in ascending order.

\paragraph{Projected moments.}
The blocks between cells are constant, and each contrast sums to zero
within its cell. Write $x_{rh}=K_{\theta_r,a}(\sqrt{A_{rh}})$ and
$y_{rh}=K_{\theta_r,a}(\sqrt{A_{rh}+6/c_r})$. Then
$G_r(\theta_r,a)v_{rhj}=\ell_{rhj}v_{rhj}$, with $\ell_{rhj}=-x_{rh}$
for $j=1,2,3$ and $\ell_{rhj}=x_{rh}-2y_{rh}$ for $j=4,5$. Premultiplication by
$v_{rhj}^{\top}$ removes the group effect, so we obtain the outcome
projection by solving one scalar equation per contrast.

We cycle through five folds as we assign cells within each group.
Using the four training folds, we fit the mean of
$v_{rhj}^{\top}Z_{rj}$ by nonlinear least squares with
$\{1+\exp(-\gamma_0-\gamma_1s_{rh})\}^{-1}$ and
$(\gamma_0,\gamma_1)\in[-8,8]^2$. If the optimizer misses the
projected-gradient tolerance, we refine the same least-squares fit.
On the same training folds we fit the transformed residual by local linear regression on
$T_{rh}$ with kernel $(1-u^2)_+$. We start with bandwidth
$H_r^{-1/5}$ and enlarge it if necessary to $1.05$ times the distance
from the evaluation point to the third nearest training point.

Let $u_{rhj}$ be the projection of
$(I-\rho G_r)Y_r-X_r\boldsymbol{\beta}_1-G_rX_r\boldsymbol{\beta}_2$
onto $v_{rhj}$. For $s=0,\ldots,4$, the $j$th moment is the sum over cells of the
product of the two residuals from centering the projected instrument
and $\ell_{rhj}^{\,s}u_{rhj}$, divided by $N=\sum_r n_r$.

\paragraph{Point estimation and intervals.}
The GMM weight is the inverse Gram matrix for the centered power
instruments at $(\vartheta,a)=(1,0)$. In computing the inverse we add
$10^{-8}$ to the diagonal on the correlation scale. We search over $\vartheta\in[0.5,2.5]$,
bound each component of $\boldsymbol{\beta}_1$ and $\boldsymbol{\beta}_2$
by $[-25,25]$, and impose
\[
 |\rho|\leq\frac{0.95}{\sqrt{1+C(\vartheta,a)^2}},
 \qquad
 C(\vartheta,a)=(1+a)e^{-0.123\vartheta}
          \left(1+5e^{-6\vartheta}+\frac{14}{e^{12\vartheta}-1}\right).
\]
The starting values for $\vartheta$ are $0.55,1,1.7,2.4$, crossed
with $a=0.15,0.85$ in the free-shape fit. The corresponding linear
GMM estimates, clipped to the bounds, start the coefficients.
We use L-BFGS-B first and \texttt{nlminb} if its projected gradient
fails the check. A bound is not grounds for discarding a fit that
passes the checks.

For a fixed shape $a$, denote the point estimate by $\check\eta$.
For interval estimation we work with powers of the eigenvalues
rescaled to $[-1,1]$. We evaluate $\hat m$ and $\hat D$ at $\check\eta$
using this basis, treating the rescaling as fixed in the derivative. With
$\hat W=\hat\Omega^{-1}$, the interval center is
\[
 \check\eta-(\hat D^\top\hat W\hat D)^{-1}\hat D^\top\hat W\hat m .
\]
In $\hat\Omega=N^{-1}\sum_{r,h}\hat\psi_{rh}\hat\psi_{rh}^{\top}$, the
cell scores include the two first-step weight derivatives. We use the
standard errors from $N^{-1}(\hat D^\top\hat W\hat D)^{-1}$ at $\check\eta$
and the 0.975 normal quantile. A failed optimization or rank check at a
given shape leaves no interval for that shape. The envelope in
Section~\ref{section: simulation} therefore uses only replications with
intervals for all nine shapes.

\paragraph{Test for no mean network spillovers.}
The null sets $\rho_0=0$ and
$\boldsymbol{\beta}_{20}=\mathbf 0$ and leaves the other values unchanged.
The seed is $180800000+100000e+10000q+m$, with $e=2$ for the null
and $e=3$ for the alternative.
We estimate the direct effects under the null by IV using the five
$s=0$ moments. At $(\vartheta,a)=(1,0)$ we then sum the remaining
moments for each instrument with weights $2^{-s}$ for $s=1,\ldots,4$.
The cell score covariance accounts for both first-step adjustments
and for estimation of the direct effects under the null. We reject
when the quadratic form exceeds the 0.95 quantile of $\chi^2_5$.

\subsection*{Verification for the Gaussian fit}
Set $a=0$ in the data-generating process and in estimation. There
are ten groups of equal size, so $H_r/N\to1/60$.
The following verification concerns the correctly specified Gaussian submodel
with $\theta_r=c_r\vartheta$ and a 12-dimensional parameter vector.
It does not establish identification for the free-mixture fit.

\paragraph{Assumption~\ref{assumption: conditional exogeneity}. Exclusion.}
Given $\mathcal F_G$, the only random term in $\varepsilon$ is
$0.35\zeta$. It is independent of the instrument draws, so
$\mathbb E[\varepsilon\mid Z,\mathcal F_G]=\mathbb E[\varepsilon\mid\mathcal F_G]$.

\paragraph{Assumption~\ref{assumption: Kernel Regularity}. Kernel regularity.}
For any $\vartheta>0$ the off-diagonal Gaussian weights determine
$d_{ik}$ and hence the cell draws. At distinct multipliers the weight ratio is a nonconstant function of
squared distance. Across cells the exponents grow at least linearly in
cell separation. The row and column sums, including those for the first
two derivatives, are thus bounded by summable series. In particular,
$|\rho_0|\sup_{\vartheta\in[0.5,2.5]} C(\vartheta,0)<0.28(1.21)<1$.

\pagebreak[3]
\paragraph{Population moments.}
For $B=c_rA_{rh}$, write $\ell_{j,\vartheta}=b_j(\vartheta)e^{-\vartheta B}$,
where $b_j(\vartheta)=-1$ for $j\le3$ and
$b_j(\vartheta)=1-2e^{-6\vartheta}$ for $j\ge4$.
Let $\ell_{j0}=\ell_{j,1}$,
$T_j=(\beta_{10,j}+\ell_{j0}\beta_{20,j})/(1-\rho_0\ell_{j0})$,
and $\tau_j=\beta_{20,j}+\rho_0\beta_{10,j}>0$. Then
\begin{equation}\label{eq:peer_gaussian_moments}
 m_{sj}(\eta)=\int \ell_{j,\vartheta}^s
       [(1-\rho\ell_{j,\vartheta})T_j-\beta_{1j}-\ell_{j,\vartheta}\beta_{2j}]\,d\nu_j .
\end{equation}
The projected instrument for $j$ has covariance
$\pi_{rj}p_{rhj}(1-p_{rhj})$ with its own projected covariate and zero
covariance with the others. Thus $\nu_j$ pools the ten laws of $B$
with weights $\pi_{rj}p_{rhj}(1-p_{rhj})/60$ and has positive density
on $(0.123,4.2)$. For a nonzero continuous function with at most three
sign changes, take a polynomial in $\ell_{j,\vartheta}$ with a root at
each sign change and choose its overall sign so that the product is
nonnegative. The product has positive integral if the function is
nonzero on the support. This rules out orthogonality to
$1,\ell_{j,\vartheta},\ell_{j,\vartheta}^2,\ell_{j,\vartheta}^3$.
By Rolle's theorem, a nonzero linear combination of four exponentials
with distinct rates has at most three zeros.

\paragraph{Assumption~\ref{assumption: rank conditions}. Rank.}
A linear dependence among the columns would give zero moments for
$d_j+e_j\ell_{j,\vartheta}+f\ell_{j,\vartheta}T_j$, with a common $f$ across $j$.
Multiplication by $1-\rho_0\ell_{j0}>0$ gives a sum of exponentials with
rates $0,1,\vartheta,1+\vartheta$. It must be identically zero by the
sign-change argument. Equating coefficients gives $d_j=0$,
$e_j=-f\beta_{10,j}$, and $f\tau_j=0$. The same equalities hold at
$\vartheta=1$ after collecting the terms with the repeated rate.
Since $\tau_j>0$, all 11 columns are linearly independent throughout
the search interval.

\paragraph{Assumption~\ref{assumption: high level separation}. Global separation.}
If $\vartheta\ne1$, clearing the denominator in
\eqref{eq:peer_gaussian_moments} gives four distinct rates. The
coefficients of $1$ and $\ell_{j0}$ are $\beta_{10,j}-\beta_{1j}$ and
$\beta_{20,j}+\rho_0\beta_{1j}$. Setting the first to zero makes the
second equal to $\tau_j>0$. The first four moments therefore exclude
$\vartheta\ne1$. At $\vartheta=1$ the coefficients are unique by
Assumption~\ref{assumption: rank conditions}.

\paragraph{Assumption~\ref{assumption: covariance shape}. Covariance shape.}
The conditional covariance of $\varepsilon$ is
$\operatorname{Var}(\varepsilon\mid\mathcal F_G)=0.35^2 I_N$.
The variance of the networked error along a cell contrast is
$\sigma_\varepsilon^2(1+\delta\ell_{j0})^2$. Almost surely there are
three distinct cell eigenvalues. They determine this quadratic, whose
constant and linear terms give $\sigma_\varepsilon^2$ and $\delta$.

\paragraph{Assumption~\ref{ass:joint-consistency}(i)--(ii). Consistency inputs.}
The moments have a unique zero on the compact search set by
Assumption~\ref{assumption: high level separation}. Independence across cells and integrable envelopes for
the moments and their derivatives give the oracle uniform law. The
limit in \eqref{eq:peer_gaussian_moments} uses $H_r/N\to1/60$.
The reference Gram matrix has a positive definite limit since a
nonzero polynomial in $\ell_{j0}$ has positive weighted squared norm.
The first-step rates below give the uniform law for the fitted moments.

\paragraph{Assumption~\ref{ass:joint-regular}(i). Interiority and local rank.}
The true parameter is in the interior of the Gaussian search set.
At $\eta_0$ the integrand for the multiplier column is
$\dot\ell_{j0}(\rho_0T_j+\beta_{20,j})$ up to sign. After multiplication by
$(1-\rho_0\ell_{j0})e^B$, this integrand and the three coefficient
integrands span $\{1,B,e^B,e^{-B}\}$ because $\tau_j\ne0$. Two
applications of Rolle's theorem bound the number of zeros of any
nonzero linear combination by three. The first four powers give rank 12.
Smooth response functions and the first-step derivative rates give
uniform convergence of the Jacobian.

\paragraph{Assumption~\ref{ass:joint-regular}(ii). Score limit and first steps.}
The oracle scores are independent across cells. Their entries are
$\ell_{j0}^s z_{rhj}\,0.35(1+\delta_0\ell_{j0})v_{rhj}^{\top}\zeta_r$.
Their fourth moments are bounded, giving the central limit theorem.
For a nonzero linear combination of the scores, at least one instrument
has a nonzero polynomial in $\ell_{j0}$. That polynomial is nonzero with
positive probability since $\ell_{j0}$ varies continuously. There is no
cancellation across instruments by conditional independence.

The instrument regression has one population minimum and a nonsingular
Hessian. Its error is $O_p(H_r^{-1/2})$. In the local linear step take
$h=H_r^{-1/5}$. The conditional means and their parameter derivatives
have bounded curvature, and a training window contains $O(H_rh)$ cells.
The resulting bias and stochastic terms bound the error by
$O_p(h^2+\sqrt{\log H_r/(H_rh)})$. Conditional on the training folds,
each linear term in the nuisance errors has mean zero and its variance
after multiplication by $\sqrt N$ is bounded by the squared error of
that nuisance fit. Multiplying the two rates gives $o_p(N^{-1/2})$
for the product term.

\paragraph{Assumptions~\ref{ass:joint-consistency}(iii) and \ref{ass:joint-regular}(iii). Optimization.}
For an exact minimizer, \ref{ass:joint-consistency}(iii) follows from its definition. With
probability tending to one the estimate is interior by consistency,
giving the first-order condition in \ref{ass:joint-regular}(iii). These are statements about
the minimizer. The stopping rules used in the experiment do not
establish the two rates or regularity of the training fits.

\paragraph{Cell covariance and one-step intervals.}
The oracle covariance obeys a law of large numbers across cells. To
compare it with the fitted covariance, consider the difference in
scores in empirical root mean square. The first instrument correction
multiplies a centered evaluation sum by the inverse training Hessian,
of order $H_r^{-1}$. For the local linear correction, one training cell
enters order $H_rh$ evaluation fits, each with weight of order
$(H_rh)^{-1}$. Conditional concentration gives a uniform square bound of
order $(\log H_r)^2/(H_rh)$. The additional error from the estimated
instrument mean is negligible, so the score difference tends to zero.
The one-step remainder is $o_p(1)$ for a root-$N$ consistent initial fit.

\paragraph{Assumption~\ref{ass:restricted-niv}. Test for no mean network spillovers.}
The null own-instrument matrix is diagonal, with entries
$\nu_j(\mathbb R)>0$. Let $Q(x)=\sum_{s=1}^4(x/2)^s$. Fitting the direct
effects subtracts the $\nu_j$-mean from $Q(\ell_{j0})$. The centered polynomial
still varies over the support. Its score has positive definite covariance,
giving the null $\chi^2_5$ limit. Here $T_j$ is the reduced-form response
under the generating kernel. The alternative gives a mean contrast
proportional to $\operatorname{Cov}_{\nu_j/\nu_j(\mathbb R)}(Q(\ell_{j0}),T_j)$.
On each contrast the two generating eigenvalues are monotone in
$B$ in the same direction. Since $Q$ is increasing on $(-1,1)$,
$Q(\ell_{j0})$ and $T_j$ move together and the covariance is positive.
Independent covariate or error shocks keep the variance positive.

\clearpage
\section{Discussion of the Span Restriction}\label{appendix: gs sufficient}

This appendix discusses Assumption~\ref{assumption: global separation universal}.
For $\Vtheta\in\Theta$, write
$$
M(\Vtheta)=\big[A_{\Vtheta},B_{\Vtheta}(\Vtheta)\big],
\qquad
v(\Vtheta)=B_{\Vtheta}(\Vtheta_0)\Vgamma_0,
$$
and let
$$
\mathcal R(\Vtheta)=\big[M(\Vtheta),v(\Vtheta)\big].
$$
The matrix $M(\Vtheta)$ has $2k+1$ columns and $\mathcal R(\Vtheta)$ has $2k+2$ columns. For each set $I$ of $2k+2$ rows, let $\mathcal R_I(\Vtheta)$ be the corresponding square submatrix and define
$$
\Psi(\Vtheta)=\sum_{|I|=2k+2}\det\{\mathcal R_I(\Vtheta)\}^2.
$$
The sum is well defined because $l\geq2k+2$. Under Assumption \ref{assumption: rank conditions}(ii),
$$
v(\Vtheta)\in\operatorname{span}M(\Vtheta)
\quad\Longleftrightarrow\quad
\operatorname{rank}\mathcal R(\Vtheta)=2k+1
\quad\Longleftrightarrow\quad
\Psi(\Vtheta)=0.
$$

\begin{assumption}[Analytic nondegeneracy]\label{assumption: analytic rank condition}
Let $\Theta\subset\RR^{d_{\theta}}$ and suppose that $\Theta$ is contained in a connected open set $U\subset\RR^{d_{\theta}}$. The following conditions hold:
\begin{enumerate}[(i)]
\item Every entry of $M(\Vtheta)$ and $v(\Vtheta)$ has a real analytic extension to $U$.
\item $\mathcal R(\Vtheta^*)$ has rank $2k+2$ for some $\Vtheta^*\in U$.
\end{enumerate}
\end{assumption}

The last condition requires one nonzero $(2k+2)\times(2k+2)$ minor at one parameter value. Economically, it rules out designs where the instruments, covariates, and network powers are too collinear.

\begin{proposition}[Generic separation]\label{proposition: analytic generic separation}
Suppose $l\geq2k+2$ and Assumptions \ref{assumption: rank conditions}(ii) and \ref{assumption: analytic rank condition} hold. Define
$$
\mathcal F_{\rm eq}
=\{\Vtheta\in\Theta\setminus\{\Vtheta_0\}:
v(\Vtheta)\in\operatorname{span}M(\Vtheta)\}.
$$
Then $\mathcal F_{\rm eq}$ has $d_{\theta}$-dimensional Lebesgue measure zero and empty interior. If $d_{\theta}=1$, $\mathcal F_{\rm eq}\cap\mathcal K$ is finite for every compact $\mathcal K\subset U$. Moreover, for every compact $\mathcal K\subset\Theta\setminus(\{\Vtheta_0\}\cup\mathcal F_{\rm eq})$,
$$
\inf_{\Vtheta\in\mathcal K}
\operatorname{dist}\{v(\Vtheta),\operatorname{span}M(\Vtheta)\}>0.
$$
\end{proposition}

\begin{proof}
Assumption \ref{assumption: analytic rank condition}(i) implies that $\Psi$ is real analytic on $U$. Part (ii) implies $\Psi(\Vtheta^*)>0$, so $\Psi$ is not identically zero. The zero set of a nonzero real analytic function on a connected open subset of $\RR^{d_{\theta}}$ has Lebesgue measure zero \citep{Mityagin2020}. Because $\mathcal F_{\rm eq}\subset\{\Vtheta\in U:\Psi(\Vtheta)=0\}$, the first conclusion follows. A set of Lebesgue measure zero cannot contain a nonempty open set, which proves the empty-interior conclusion.

When $d_{\theta}=1$, the zeros of a nonzero real analytic function are isolated. An infinite subset of a compact $\mathcal K\subset U$ has an accumulation point in $\mathcal K$, so $\Psi$ can have only finitely many zeros in $\mathcal K$.

For the last conclusion, let
$$
P(\Vtheta)=M(\Vtheta)
\{M(\Vtheta)^{\top}M(\Vtheta)\}^{-1}M(\Vtheta)^{\top}.
$$
Assumption \ref{assumption: rank conditions}(ii) makes $P(\Vtheta)$ continuous on $\Theta$. Hence
$$
\Delta(\Vtheta)=\|(I_l-P(\Vtheta))v(\Vtheta)\|
$$
is continuous. It is strictly positive on $\mathcal K$ by the definition of $\mathcal F_{\rm eq}$ and therefore has a strictly positive minimum on $\mathcal K$.
\end{proof}

Assumption \ref{assumption: global separation universal} is equivalent to $\mathcal F_{\rm eq}=\varnothing$. Proposition \ref{proposition: analytic generic separation} shows that, under analytic nondegeneracy, the false candidates violating its inequality form a null set with empty interior. The one-point rank check does not exclude isolated false roots.

The Gaussian and logistic kernels
$$
K^{\mathrm G}_{\theta}(d)=\exp(-\theta d),
\qquad
K^{\mathrm L}_{\theta}(d)=\frac{2\exp(-\theta d)}{1+\exp(-\theta d)}
$$
are real analytic in $\theta$ for every finite $d$. Assumption \ref{assumption: Kernel Regularity}(ii) also permits a common version of $\mu_{G(\Vtheta)}$, because $\sigma\{G(\Vtheta)\}$ is the same sigma-field for every $\Vtheta$. Matrix multiplication and finite network powers preserve analyticity. Thus Assumption \ref{assumption: analytic rank condition}(i) follows, for example, when distances are bounded, the data have the required finite moments, and the local power series of the moment integrands admit an integrable envelope. Kernel analyticity does not imply Assumption \ref{assumption: rank conditions}(ii) or the one-point rank condition; those are relevance conditions on the joint distribution of instruments, covariates, and outcomes.

\clearpage
\bibliographystyle{chicago}
\bibliography{refs_v1}

\end{document}